\documentclass[11pt,draftcls,onecolumn]{IEEEtran}
\IEEEoverridecommandlockouts                              
\usepackage{amsmath}
\usepackage{color}
\usepackage{graphicx}
\usepackage{latexsym}

\usepackage{epsfig} 

\title{Robust Recursive State Estimation with Random Measurements Droppings}

\author{Tong Zhou
\thanks{This work was supported in part by the
973 Program under Grant 2009CB320602, the National Natural Science
Foundation of China under Grant 61174122 and 61021063, and the
Specialized Research Fund for the Doctoral Program of Higher
Education, P.R.C., under Grant 20110002110045.}
\thanks{T.Zhou is with the Department of Automation and TNList, Tsinghua University, Beijing, 100084,
CHINA. (Tel: 86-10-62797430; Fax: 86-10-62786911; e-mail:
tzhou@mail.tsinghua.edu.cn.)} }

\begin{document}
\renewcommand{\thefootnote}{\fnsymbol{footnote}}
\maketitle 
\renewcommand{\thepage}{36--\arabic{page}}
\setcounter{page}{1}

\begin{abstract}
A recursive state estimation procedure is derived for a linear time
varying system with both parametric uncertainties and stochastic
measurement droppings. This estimator has a similar form as that of
the Kalman filter with intermittent observations, but its parameters
should be adjusted when a plant output measurement arrives. A new
recursive form is derived for the pseudo-covariance matrix of
estimation errors, which plays important roles in analyzing its
asymptotic properties. Based on a Riemannian metric for positive
definite matrices, some necessary and sufficient conditions have
been obtained for the strict contractiveness of an iteration of this
recursion. It has also been proved that under some controllability
and observability conditions, as well as some weak requirements on
measurement arrival probability, the gain matrix of this recursive
robust state estimator converges in probability one to a stationary
distribution. Numerical simulation results show that estimation
accuracy of the suggested procedure is more robust against
parametric modelling errors than the Kalman filter.

\end{abstract}

\IEEEpeerreviewmaketitle

\section{Introduction}

State estimation is one of the essential issues in systems and
control theory, and has attracted extensive attentions from various
fields for a long time. Major cornerstones in this field include the
Winner filter, the Kalman filter, the particle filter, the
set-membership filter, etc. While the developed state estimators
have numerous distinguished forms in their appearances, most of them
are in essence closely related to least squares estimations, and some
of them can even be regarded as its extensions to various different
situations, such as multiple-input multiple-output systems, systems
disturbed by non-normal external noises, etc.
\cite{cortes09,dfg01,gvz00,kalman60,ksh00,ln12,yl09}.

With recent significant advancements of network technologies,
utility of wireless networks, internet, etc., is strongly expected
in increasing structure flexibilities and reducing infrastructure
investments in building a large scale system, and/or implementing
remote monitoring, etc. To make this conception applicable to actual
engineering problems, however, various new theoretical challenges
should be attacked. For example, in a communication network, data
packets carrying an observed plant output can be randomly lost,
delayed or even their original order can be changed, due to traffic
conditions of the internet and/or propagation property variations of
wireless medium, etc.\cite{ssfpjs04,km12,ln12,mn12}.

Over the last decade, various efforts have been devoted to state
estimations with random missing measurements. In \cite{ssfpjs04}, it
is proved that when a plant model is accurate and external
disturbances are normally distributed, the Kalman filter is still
optimal in the sense of mean squared errors (MSE) even if there
exist random measurement droppings, provided that information is
available on whether or not the received data is a measured plant
output. It has also been proved there that for an
unstable plant, even it is both controllable and observable, the
expectation of the covariance matrix of estimation errors may become
infinitely large when the probability of receiving a plant output
measurement is too low. Afterwards, it has been argued by many
researchers that it may be more appropriate to investigate the
probability distribution of this covariance matrix, as events of
very low probability may cause an infinite expectation.
Particularly, some upper and lower bounds have been derived in
\cite{sem10,rmf11} for the probability of this covariance matrix
being smaller than a prescribed positive definite matrix (PDM). In
\cite{ms12}, it is proved that under some controllability and
observability conditions, the trace of this covariance matrix
follows a power decay law for an unstable plant with a
diagonalizable state transition matrix. On the basis of the
contractiveness of Riccati recursions and convergence of random
iterated functions, it has been proved in \cite{Censi11} that this
covariance matrix usually converges to a stationary distribution
that is independent of the plant initial states, no matter the
communication channel is described by a Bernoulli process, a Markov
chain or a semi-Markov chain. In \cite{km12}, it is proved that when
the observation arrival is modeled by a Bernoulli process and the
packet arrival probability approaches to 1, the covariance matrix
converges weakly to a unique invariant distribution that satisfies a
moderate deviation principle with a good rate function. In
\cite{wyhl05,mn12}, one-step prediction is investigated using an
estimator with a prescribed structure that tolerates both random
measurement droppings and some specific kinds of parametric
modelling errors, and a recursive estimation procedure has been
respectively derived through minimizing an upper bound of the
covariance matrix of estimation errors. While the obtained
estimators share a similar form as that of the Kalman filter, a
parameter should be adjusted on-line to guarantee the existence of
the inverse of a matrix, which may restrict successful
implementation of the developed recursive estimation procedure.

These investigations have clarified many important characteristics
about state estimations with random measurement arrivals, and have
greatly advanced studies on analysis and synthesis of networked
systems. But except \cite{wyhl05,mn12}, plant models are assumed
precisely known in almost all these investigations. In actual
engineering applications, however, model errors, which include
parametric deviations from nominal values, unmodelled dynamics,
approximation errors due to plant nonlinear dynamics, etc., are
usually unavoidable. In addition, it has also been widely observed
that estimation accuracies of some optimal estimators, including the
Kalman filter, may be deteriorated appreciably by modelling errors
\cite{ksh00,simon06,dfg01,george12,gvz00,nbt07,yl09,zhou10b}.

To make a state estimator robust against modelling errors, various
approaches have been proposed, such as the $H_{\infty}$ norm
optimization based method, the guaranteed cost based approach, etc.
Among these approaches, the sensitivity penalization based method
has some appreciated properties, such as its similarities to the
Kalman filter in estimation procedures, no requirements on
verification of matrix inequalities during estimate updates,
capability of dealing with various kinds of parametric modelling
errors, etc. \cite{zhou10b,zhou11}. In \cite{lz11}, an attempt has
been made to extend this method to situations in which random
measurement dropping tolerances are required. While some results
have been obtained, its success is rather limited, noting that the
developed estimation algorithm requires some ergodic conditions on
the received signal which can hardly be satisfied by a time varying
system. In addition, the developed estimation procedure has not
efficiently utilized the information contained in a received signal
about whether or not it is the measurement of a plant output.
Another restriction of the results in \cite{lz11} is that they are
only valid for systems with a communication channel described by the
Bernoulli random process.

In this paper, we reinvestigate the extension of the sensitivity
penalization based robust state estimation method to systems with
random measurement droppings. All the above limitations have been
successfully removed. Through introducing a new cost function, a
novel recursive procedure has been derived for state estimation with
random missing measurements. This procedure also reduces to the
Kalman filter when the plant model is accurate. A new recursion
formula has been established for the pseudo-covariance matrix (PCM)
of estimation errors which makes it possible to analyze asymptotic
properties of the developed robust state estimator (RSE). It has
also been proved that under some controllability and observability
conditions on the nominal and adjusted system matrices, as well as
some weak requirements on the random measurement loss process, the
gain matrix of the RSE converges with probability one to a
stationary distribution that is independent of its initial values.
Some numerical simulation results are also provided to illustrate
its characteristics in estimating states of a plant with both
parametric modelling errors and random measurement droppings.

The outline of this paper is as follows. At first, in Section II,
the problem formulation is provided and the estimation procedure is
derived. Afterwards, some related properties on Riccati recursions
are introduced in Subsection III.A as preliminary results, while
asymptotic characteristics of the estimator are investigated in
Subsection III.B. A numerical example is then provided in Section IV
to illustrate the effectiveness of the proposed estimator. Finally,
some concluding remarks are given in Section V summarizing
characteristics of the suggested method. An appendix is included to
give proofs of some technical results.

The following notation and symbols are adopted.
$||\cdot||$ stands for the Euclidean norm of a
vector, while $||x||_{W}$ is a shorthand for $\sqrt{x^{T}Wx}$.
${\rm\bf diag}\!\{X_{i}|_{i=1}^{L}\}$ denotes a block diagonal
matrix with its $i$-th diagonal block being $X_{i}$, while ${\rm\bf
col}\!\{X_{i}|_{i=1}^{L}\}$ the vector/matrix stacked by
$X_{i}|_{i=1}^{L}$ with its $i$-th row block vector/matrix being
$X_{i}$. $\left[X_{ij}|_{i=1,j=1}^{i=M,j=N}\right]$ represents a
matrix with $M\times N$ blocks and its $i$-th row $j$-th column
block matrix being $X_{ij}$, while the product $\Phi_{k1}\Phi_{k1-1\;{\rm or}\; k1+1}\cdots\Phi_{k2}$ is denoted by
$\prod_{j=k1}^{k2}\Phi_{j}$. The superscript $T$ is used to denote
the transpose of a matrix/vector, and $X^{T}WX$ or $XWX^{T}$ is
sometimes abbreviated as $(\star)^{T}WX$ or $XW(\star)^{T}$,
especially when the term $X$ has a complicated expression. ${\rm\bf
D}_{et}\!\{\star\}$ stands for the determinant of a matrix, while
${\rm\bf L}_{ip}\!\{\star\}$ the Lipschitz constant of a function.
${\rm\bf P}_{r}(\cdot)$ is used to denote the probability of the
occurrence of a random event, while ${\rm\bf
E}_{\{\sharp\}}\!\{\star\}$ the mathematical expectation of a matrix
valued function (MVF) $\star$ with respect to the random variable
$\sharp$. The subscript $\sharp$ is usually omitted when it is
obvious.

\section{The Robust State Estimation Procedure}

Consider a linear time varying dynamic system $\rm\bf\Sigma$ with
both parametric modelling errors due to imperfect information about
the plant dynamics and stochastic measurement loss due to
communication failures. Assume that its input output relations can
be described by the following discrete state-space model,
\begin{equation}
{\rm\bf\Sigma}: \hspace{0.5cm}\left\{\begin{array}{l}
x_{t+1}=A_{t}(\varepsilon_{t})x_{t}+B_{t}(\varepsilon_{t})w_{t} \\
y_{t}=\gamma_{t}C_{t}(\varepsilon_{t})x_{t}+v_{t} \end{array}
\right. \label{eqn:1}
\end{equation}
Here, $\varepsilon_{t}$ is a $n_{e}$ dimensional vector representing
parametric errors of the plant state-space model at the time instant
$t$, $\gamma_{t}$ is a random variable characterizing successes and
failures of communications between the plant output measurement
sensors and the state estimator. It takes the value of $1$ when a
plant output measurement is successfully transmitted, and the value
of $0$ when the communication channel is out of order. Vectors
$w_{t}$ and $v_{t}$ denote respectively process noises and composite
influences of measurement errors and communication errors. It is
assumed in this paper that both $w_{t}$ and $v_{t}$ are white and
normally distributed, ${\rm\bf E}\!\left({\rm\bf
col}\!\{w_{t},v_{t},x_{0}\}\right)=0$ and ${\rm\bf E}\!\left({\rm\bf
col}\!\{w_{t}, v_{t},x_{0}\}{\rm\bf col}^{T}\!\{w_{s},
v_{s},x_{0}\}\right)={\rm\bf diag}\{Q_{t}\delta_{ts},
R_{t}\delta_{ts},P_{0}\}$, $\forall t,s>0$. Here, $\delta_{ts}$
stands for the Kronecker delta function, and
$Q_{t}$ and $R_{t}$ are known positive definite MVFs of the temporal
variable $t$, while $P_{0}$ is a known PDM. These assumptions imply
that these two external disturbances are independent of each other,
and are also independent of the plant initial conditions. Another
hypothesis adopted in this paper is that all the system matrices
$A_{t}(\varepsilon_{t})$, $B_{t}(\varepsilon_{t})$ and
$C_{t}(\varepsilon_{t})$  are time varying but
known MVFs with all elements differentiable with respect to every
element of $\varepsilon_{t}$ at each time instant. It is also
assumed throughout this paper that the state vector $x_{t}$ of the
dynamic system $\rm\bf\Sigma$ has a dimension $n$, and an indicator
is included in the received signal $y_{t}$ that reveals whether or
not it contains information about plant outputs.

In the above descriptions,
$A_{t}(\varepsilon_{t})$, $B_{t}(\varepsilon_{t})$ and
$C_{t}(\varepsilon_{t})$ with $\varepsilon_{t}=0$ are plant nominal
system matrices. According to the adopted hypotheses, all these
matrices are assumed known. The vector $\varepsilon_{t}$ stands for
deviations of plant actual parameters from their nominal values,
which are permitted to be time varying and are generally unknown. In
model based robust system designs or state estimations, however,
some upper magnitude bounds or stochastic properties are usually
assumed available for this parametric error vector
\cite{gvz00,george12,kimura97,ksh00,zhou10b}. While this kind of
information is important in determining the design parameter
$\mu_{t}$ of the following Equation (\ref{eqn:7}), which is also
illustrated by the numerical example of Section IV, it is not used
in this paper.

The main objectives of this paper are to derive an estimate for the
plant state vector $x_{t}$ using the received plant output
measurements $y_{i}|_{i=0}^{t}$ and information about the
corresponding realization of $\gamma_{i}|_{i=0}^{t}$, as well as to
analyze its asymptotic statistical characteristics.

When the plant state space model for a linear time varying system is
precise, a widely adopted state estimation procedure is the Kalman
filter, which can be recursively realized and have achieved
extensive success in actual engineering applications
\cite{kalman60}. This estimation procedure, however, may sometimes
not work very satisfactorily due to modelling errors. To overcome
this disadvantage, various modifications have been suggested which
make the corresponding estimation accuracy more robust against
modelling errors \cite{george12,
simon06,ksh00,mn12,wyhl05,yl09,zhou10b}. Among these modifications,
one effective method is based on sensitivity penalization, in which
a cost function is constructed on the basis of
least squares/likelihood maximization
interpretations for the Kalman filter and a penalization on the
sensitivity of its innovation process to modelling errors
\cite{zhou10b,zhou11}.

More precisely, assume that plant parameters are
accurately known for the above dynamic system $\rm\bf\Sigma$ and
there do not exist measurement droppings. These requirements are
respectively equivalent to $\varepsilon_{t}\equiv 0$ and
$\gamma_{t}\equiv 1$. Let $\hat{x}_{t|t}^{[kal]}$ and
$P_{t|t}^{[kal]}$ represent respectively the estimate of the Kalman
filter for the plant state vector $x_{t}$ based on plant output
measurements $y_{i}|_{i=0}^{t}$ and the covariance matrix of the
corresponding estimation errors. Then, $\hat{x}_{t+1|t+1}^{[kal]}$,
the estimate of the plant state vector at the time instant $t+1$
based on plant output measurements $y_{i}|_{i=0}^{t+1}$, can also be
recursively expressed as
$\hat{x}_{t+1|t+1}^{[kal]}=A_{t}(0)\hat{x}_{t|t+1}^{[kal]}+B_{t}(0)\hat{w}_{t|t+1}^{[kal]}$,
in which $\hat{x}_{t|t+1}^{[kal]}$ and $\hat{w}_{t|t+1}^{[kal]}$
stand for vectors $x_{t|t+1}$ and $w_{t|t+1}$ that minimize the cost
function
$J^{[kal]}(x_{t|t+1},\;w_{t|t+1})=||x_{t|t+1}-\hat{x}_{t|t}^{[kal]}||_{(P_{t|t}^{[kal]})^{-1}}^{2}+||w_{t|t+1}||_{Q_{t}^{-1}}^{2}
+||e_{t}(0,\;0)||_{R_{t+1}^{-1}}^{2}$, in which
$e_{t}(\varepsilon_{t},\;\varepsilon_{t+1})=y_{t+1}-C_{t+1}(\varepsilon_{t+1})[A_{t}(\varepsilon_{t})x_{t|t+1}+B_{t}(\varepsilon_{t})w_{t|t+1}]$
that is generally called the innovation process in estimation theory
when the plant model is accurate \cite{ksh00,simon06}. Note that
from the Markov properties of the plant dynamics and the fact that
the Kalman filter is a linear function of plant output measurements,
it can be claimed that both the plant state vector and its Kalman
filter based estimate are normally distributed. Based on these
facts, it can be further declared that the aforementioned
$\hat{x}_{t|t+1}^{[kal]}$ and $\hat{w}_{t|t+1}^{[kal]}$ are in fact
respectively the $y_{i}|_{i=0}^{t+1}$ based maximum likelihood
estimates of $x_{t}$ and $w_{t}$. On the other hand, from the
expression of the cost function $J^{[kal]}(x_{t|t+1},\;w_{t|t+1})$,
the Kalman filter can also be interpreted as a least squares
estimator \cite{ksh00,simon06}.

When $\varepsilon_{t}\not\equiv 0$ and {\it only}
nominal plant parameters are known, in order to increase robustness
of the Kalman filter against parametric modelling errors, it is
suggested in \cite{zhou10b} to add some penalties on the sensitivity
of the innovation process
$e_{t}(\varepsilon_{t},\;\varepsilon_{t+1})$ to modelling errors
into this cost function. The rationale is that deviations of this
innovation process from its nominal values reflect contributions of
parametric modelling errors to prediction errors of the Kalman
filter about plant outputs. Note that when
$\varepsilon_{i}\not\equiv 0$,
$e_{t}(\varepsilon_{t},\;\varepsilon_{t+1})$ is the only factor in
the cost function $J^{[kal]}(x_{t|t+1},\;w_{t|t+1})$ that depends on
system parameters. This means that reduction of its deviations due
to modelling errors in fact also reduces the counterpart of this
cost function, and therefore increases robustness of the
corresponding state estimator. Noting also that accurate expression
for this deviation generally has a complicated form and may make the
corresponding estimation problem mathematically intractable, it is
suggested in \cite{zhou10b} to consider its first order
approximation, that is, to linearize
$e_{t}(\varepsilon_{t},\;\varepsilon_{t+1})$ at the origin.
Specifically, the cost function $J^{[kal]}(x_{t|t+1},\;w_{t|t+1})$
is modified to
\begin{displaymath}
J^{[sen]}(x_{t|t+1},w_{t|t+1})\!=\!\mu_{t}J^{[kal]}(x_{t|t+1},w_{t|t+1})\!+\!(1\!-\!\mu_{t})\!\!\left.\sum_{k=1}^{n_{e}}\!\!\left(
\left|\left|\frac{\partial
e_{t}(\varepsilon_{t},\;\varepsilon_{t+1})}{\partial
\varepsilon_{t,k}}\right|\right|_{2}^{2}\!\!+\!\!\left|\left|\frac{\partial
e_{t}(\varepsilon_{t},\;\varepsilon_{t+1})}{\partial
\varepsilon_{t+1,k}}\right|\right|_{2}^{2}\!\right)\!\right|\!\!\!{\footnotesize\begin{array}{l}
\\ \varepsilon_{t}\!=\!0 \vspace{-0.25cm}
\\ \varepsilon_{t+1}\!=\!0\end{array}}
\end{displaymath}
in which $\mu_{t}$ is a positive design parameter belonging to
$(0,\;1]$ that reflects a trade-off between nominal value of
estimation accuracy and penalization on the first order
approximation of deviations of the innovation process due to
parametric modelling errors.

Based on this modified cost function
$J^{[sen]}(x_{t|t+1},\;w_{t})$, a state estimation procedure is
derived in \cite{zhou10b}. It has also been proved there that except
some parameter adjustments, this estimation procedure has a similar
form as that of the Kalman filter, and its estimation gain matrix
also converges to a constant matrix if some controllability and
observability conditions are satisfied. Boundedness of the
covariance matrix of its estimation errors has also been established
under some weak conditions like quadratic stability
of the plant and contractiveness of the parametric errors, etc. It
has been shown that the estimation procedure reduces to the Kalman
filtering if parametric uncertainties disappear
\cite{zhou10b,zhou11}.

In this paper, the same approach is adopted to deal with the state
estimations for the linear time varying dynamic system
$\rm\bf\Sigma$ in which both parametric uncertainties and random
measurement droppings exist. It is worthwhile to point out that
although this extension has been attempted in \cite{lz11}, the
success is rather limited. One of the major restrictions on
applicability of the obtained results is the implicit ergodic
requirement on the received plant output measurements, which is
generally not satisfied by a time varying system. Another major
restriction is that in developing the estimation procedure,
information about the realization of the random process $\gamma_{t}$
has not been efficiently utilized, which makes the corresponding
estimation accuracy sometimes even worse than the traditional Kalman
filter that does not take either parametric errors or random
measurement loss into account. These disadvantages have been
successfully overcome in this paper through introducing another cost
function which is more appropriate in dealing with simultaneous
existence of parametric uncertainties and random measurement
droppings.

More precisely, assume that at the time instant $t$, an estimate is
obtained for the plant state using the received plant output
measurements $y_{i}|_{i=0}^{t}$, denote it by $\hat{x}_{t|t}$. Let
$P_{t|t}$ represent the PCM of the corresponding state estimation
errors. Construct a cost function $J(x_{t|t+1},\;w_{t|t+1})$ as
follows,
\begin{eqnarray}
J(x_{t|t+1},\;w_{t|t+1})&=&\frac{1}{2}\left\{\mu_{t}\left[||x_{t|t+1}-\hat{x}_{t|t}||_{P_{t|t}^{-1}}^{2}+||w_{t|t+1}||_{Q_{t}^{-1}}^{2}\right]
+\gamma_{t+1}\left[\mu_{t}||e_{t}(0,\;0)||_{R_{t+1}^{-1}}^{2}+(1-\mu_{t})\times\right.\right.\nonumber\\
& & \hspace*{2cm}\left.\left.\left.\sum_{k=1}^{n_{e}}\left(
\left|\left|\frac{\partial
e_{t}(\varepsilon_{t},\;\varepsilon_{t+1})}{\partial
\varepsilon_{t,k}}\right|\right|_{2}^{2}+\left|\left|\frac{\partial
e_{t}(\varepsilon_{t},\;\varepsilon_{t+1})}{\partial
\varepsilon_{t+1,k}}\right|\right|_{2}^{2}\right)\right|\!\!\!{\footnotesize\begin{array}{l}
\\ \varepsilon_{t}=0 \vspace{-0.25cm}
\\ \varepsilon_{t+1}=0\end{array}}\right]\right\}
\label{eqn:7}
\end{eqnarray}
Here, both
$e_{t}(\varepsilon_{t},\;\varepsilon_{t+1})$ and $\mu_{t}$ have the
same definitions as those in the aforementioned sensitivity
penalization based robust estimator design. While $\mu_{t}$
selection is an important issue in designing a robust state
estimator and depends on properties of parametric modelling errors
\cite{zhou10b,zhou11}, it is assumed given in this paper.

In this cost function, $\gamma_{t+1}$ is explicitly utilized which
is generally available in communications after $y_{t+1}$ is
received. In fact, to make this information accessible, the only
requirement is to include an indication code in a communication
channel which is usually possible \cite{ssfpjs04,km12,ms12}.
On the other hand, if $\gamma_{t+1}=1$, that is, if
there is no measurement loss from the system output measurement
sensor to the state estimator, this cost function is equivalent to
$J^{[sen]}(x_{t|t+1},\;w_{t|t+1})$, which means that as some new
information on $x_{t}$ contained in $y_{t+1}$ has arrived at the
time instant $t+1$, its estimate should be updated in a robust way
that is not sensitive to parametric modelling errors. If a
measurement dropping happens in communications, then, $y_{t+1}$ does
not contain any information about the plant output and therefore
$x_{t}$. In this case, as the existing estimate on $x_{t}$ is
optimal and no new information about it arrives, there is no need to
update this estimate, which is equivalent to that the cost function
does not depend on either the nominal value of
$e_{t}(\varepsilon_{t},\;\varepsilon_{t+1})$ or its sensitivity to
parametric modelling errors. In other words, when no plant output
measurement is available at a time instant, the estimator can only
predict the plant state vector using the previously collected
information, and this physically obvious characteristic has been
satisfactorily reflected by the above cost function. From these
aspects, it appears safe to declare that the cost function
$J(x_{t|t+1},\;w_{t|t+1})$ has simultaneously satisfied both the
optimality requirements and the robustness requirements in state
estimations under simultaneous existence of parametric modelling
errors and measurement loss, and is therefore physically more
reasonable than that of \cite{lz11}.

However, it is worthwhile to mention that in the
above cost function $J(x_{t|t+1},\;w_{t|t+1})$, the purpose to
include a penalty on the sensitivity of the innovation process
$e_{t}(\varepsilon_{t},\varepsilon_{t+1})$ to modelling errors is to
increase the robustness of state estimations against deviations of
plant parameters from their nominal values. There are also many
important practical situations, for example, fault detection, signal
segmentation, financial market monitoring, etc., in which an
estimate sensitive to actual parameter variations are more greatly
appreciated \cite{bn93}. Under these situations, the above cost
function, and therefore the corresponding state estimate procedure,
are no longer appropriate.

Let $\hat{x}_{t|t+1}$ and $\hat{w}_{t|t+1}$ denote the optimal
$x_{t|t+1}$ and $w_{t|t+1}$ that minimize the above cost function
$J(x_{t|t+1},w_{t|t+1})$. Then, according to the sensitivity
penalization approach towards robust state estimations, an
$y_{i}|_{i=0}^{t+1}$ based estimate of the plant state vector
$x_{t+1}$, denote it by $\hat{x}_{t+1|t+1}$, can be constructed as
follows,
\begin{equation}
\hat{x}_{t+1|t+1}=A_{t}(0)\hat{x}_{t|t+1}+B_{t}(0)\hat{w}_{t|t+1}
\label{eqn:8}
\end{equation}

When there are no parametric uncertainties in the plant model, the
matrix $P_{t|t}$ is in fact the covariance matrix of the estimation
errors of the Kalman filter. This makes it possible to explain
$\hat{x}_{t|t+1}$ and $\hat{w}_{t|t+1}$ respectively as the
$y_{i}|_{i=0}^{t+1}$ based maximum likelihood estimates of $x_{t}$
and $w_{t}$ \cite{ksh00,simon06, zhou10b}. But when there exist
modelling errors in the system matrices $A_{t}(\varepsilon_{t})$,
$B_{t}(\varepsilon_{t})$ and $C_{t}(\varepsilon_{t})$, physical
interpretations of the matrix $P_{t|t}$ need further clarifications
\cite{zhou10b}. To avoid possible misunderstandings, it is called
pseudo-covariance matrix (PCM) in this paper.

Based on the above construction procedure, a recursive estimation
algorithm can be derived for the state vector of a plant with both
parametric uncertainties and random measurement loss, while its
proof is deferred to the appendix.

\hspace*{-0.4cm}{\bf Theorem 1.} Let $\lambda_{t}$ denote
$\frac{1-\mu_{t}}{\mu_{t}}$. Assume that both $P_{t|t}$ and $Q_{t}$
are invertible. Then, the estimate of the state vector $x_{t+1}$ of
the dynamic system $\rm\bf\Sigma$ based on $y_{k}|_{k=0}^{t+1}$ and
Equations (\ref{eqn:7}) and (\ref{eqn:8}) has the following
recursive expression,
\begin{equation}
\hat{x}_{t+1|t+1}=\left\{\begin{array}{ll} A_{t}(0)\hat{x}_{t|t} &
\gamma_{t+1}=0 \\
\hat{A}_{t}(0)\hat{x}_{t|t}+P_{t+1|t+1}C_{t+1}^{T}(0)R_{t+1}^{-1}\{y_{t+1}-C_{t+1}(0)\hat{A}_{t}(0)\hat{x}_{t|t}\}
& \gamma_{t+1}=1
\end{array}\right.
\end{equation}
Moreover, the PCM $P_{t|t}$ can be recursively updated as
\begin{equation}
P_{t+1|t+1}\!=\!\left\{\!\!\begin{array}{ll}
A_{t}(0)P_{t|t}A_{t}^{T}(0)+B_{t}(0)Q_{t}B_{t}^{T}(0) &
\gamma_{t+1}=0 \\
\left\{\left[A_{t}(0)\hat{P}_{t|t}A_{t}^{T}(0)+\hat{B}_{t}(0)\hat{Q}_{t}\hat{B}_{t}^{T}(0)\right]^{-1}+C_{t+1}^{T}(0)R_{t+1}^{-1}C_{t+1}(0)\right\}^{-1}
& \gamma_{t+1}=1
\end{array}\right.
\end{equation}
in which
\begin{eqnarray*}
& & \hat{P}_{t|t}=(P_{t|t}^{-1}+\lambda_{t}S_{t}^{T}S_{t})^{-1}, \hspace{0.5cm}
\hat{Q}_{t}=\left[Q_{t}^{-1}+\lambda_{t}T_{t}^{T}(I+\lambda_{t}S_{t}P_{t|t}S_{t}^{T})T_{t}\right]^{-1} \\
& &
\hat{B}_{t}(0)=B_{t}(0)-\lambda_{t}A_{t}(0)\hat{P}_{t|t}S_{t}^{T}T_{t},\hspace{0.5cm}
\hat{A}_{t}(0)=[A_{t}(0)-\hat{B}_{t}(0)\hat{Q}_{t}T_{t}^{T}S_{t}][I-\lambda_{t}\hat{P}_{t|t}S_{t}^{T}S_{t}]\\
& & S_{t}={\rm\bf col}\!\left.\left\{\left[\begin{array}{cc}
C_{t+1}(\varepsilon_{t+1})\frac{\partial(A_{t}(\varepsilon_{t}))}{\partial\varepsilon_{t,k}}
\\
\frac{\partial(C_{t+1}(\varepsilon_{t+1}))}{\partial\varepsilon_{t+1,k}}A_{t}(\varepsilon_{t})
\end{array}\right]_{k=1}^{n_{e}}\right\}\right|\!\!\!{\footnotesize\begin{array}{l}
\\ \varepsilon_{t}=0 \vspace{-0.25cm}
\\ \varepsilon_{t+1}=0\end{array}}\!\!\!\!\!,\hspace{0.25cm}
T_{t}={\rm\bf col}\!\left.\left\{\left[\begin{array}{cc}
C_{t+1}(\varepsilon_{t+1})\frac{\partial(B_{t}(\varepsilon_{t}))}{\partial\varepsilon_{t,k}}
\\
\frac{\partial(C_{t+1}(\varepsilon_{t+1}))}{\partial\varepsilon_{t+1,k}}B_{t}(\varepsilon_{t})
\end{array}\right]_{k=1}^{n_{e}}\right\}\right|\!\!\!{\footnotesize\begin{array}{l}
\\ \varepsilon_{t}=0 \vspace{-0.25cm}
\\ \varepsilon_{t+1}=0\end{array}}
\end{eqnarray*}

Note that when $\gamma_{t+1}=0$, the above estimator is just a
one-step state predictor using nominal system matrices. On the other
hand, when $\gamma_{t+1}=1$, the above estimator still has the same
structure as that of the Kalman filter, except that the nominal
system matrices $A_{t}(0)$, $B_{t}(0)$, etc., should be adjusted to
reduce sensitivity of estimation accuracy to modelling errors. The
adjustment method of these matrices is completely the same as that
of the sensitivity penalization based RSE developed in
\cite{zhou10b} and is no longer required if the design parameter
$\mu_{t}$ is selected to be $1$. This means that the above recursive
estimation procedure is consistent with both RSE of \cite{zhou10b}
and the Kalman filtering with intermittent observations (KFIO)
reported in \cite{ssfpjs04}. As a by-product of this investigation,
another derivation of KFIO is obtained, in which the assumption is
no longer required that the covariance matrix of measurement noise
tends to infinity when a measured plant output is lost by a
communication channel. This assumption is essential in the KFIO
derivations given in \cite{ssfpjs04}, but does not appear very
natural from an engineering point of view.

However, Theorem 1 also makes it clear that when there exist both
parametric modelling errors and random measurement droppings, the
system matrices used by the estimator depend on whether or not
$y_{t+1}$ contains information about plant outputs. This makes the
estimator different from KFIO, and also makes analysis more
mathematically involved about its asymptotic characteristics.

\section{Convergence Analysis of the Robust State Estimator}

In evaluating performances of a state estimator, one extensively
utilized metric is about its convergence. A general belief is that
if an estimator does not converge, satisfactory performance can not
be anticipated. It is now well known that for a linear time
invariant system, under some controllability and observability
conditions, the gain of the Kalman filter converges to a constant
matrix. This property makes it possible to approximate the Kalman
filter satisfactorily with an a constant gain observer
\cite{simon06,ksh00}.

When plant output measurements are randomly received, $\gamma_{t}$
of Equation (\ref{eqn:1}) is a random process. This makes the PCM
$P_{t|t}$, and therefore the gain matrix of the state estimator,
also a random process. Generally, it can not be
anticipated that they converge to constant matrices, but it is still
theoretically and practically interesting to see whether or not they
have stationary distributions \cite{Censi11,km12}. Note that both
the matrix $C_{t}(0)$ and the matrix $R_{t}$ are deterministic MVFs
of the temporal variable $t$. An interesting and basic issue here
is therefore that whether or not the matrix $P_{t|t}$ converges to a
stationary distribution.

Although the derived RSE has a similar structure as that of KFIO,
the recursions for the PCM $P_{t|t}$ have a more complicated form,
as system matrices should be adjusted when a received packet
contains information about plant output. This adjustment invalidates
the relatively simple relations between the $P_{t|t-1}$s of KFIO
with respectively $\gamma_{t}=1$ and $\gamma_{t}=0$ that play
essential roles in establishing its asymptotic properties
\cite{Censi11,km12,ms12}. As a matter of fact, this adjustment makes
the corresponding analysis much more mathematically involved for the
RSE developed in this paper, which is abbreviated for brevity to
RSEIO in the rest of this paper, and leads to conclusions different
from those of KFIO.

\subsection{Preliminary Results for Convergence Analysis}

To investigate the asymptotic properties of RSEIO, some preliminary
results are required, which include a matrix transformation, a
Riemannian distance for PDMs and some characteristics of a
Hamiltonian matrix. Some of them have already been utilized in
analyzing asymptotic properties of KFIO and Kalman filter with
random coefficients \cite{Bougerol93,Censi11}.

Assume that $P$ and $Q$ are two $n\times n$ dimensional PDMs. Let
$\lambda_{i}$ denote the eigenvalues of the matrix $PQ^{-1}$. The
Riemannian distance between these two matrices, denote it by
$\delta(P,Q)$, is defined as $\delta(P,Q)=\sqrt{\sum_{i=1}^{n}{\rm
log}^{2}\lambda_{i}}$. An attractive property of this distance is
its invariance under conjugacy transformations and inversions. It is
now also known that when equipped with this distance, the space of
$n\times n$ dimensional PDMs is complete. This metric, although not
widely known, has been recognized very useful for many years in
studying asymptotic properties of Kalman filtering with random
system matrices \cite{Bougerol93}.  Its effectiveness in studying
asymptotic properties of KFIO has also been discovered recently
\cite{Censi11}.

For matrices $P$ and
$\Phi=\left[\left.\Phi_{ij}\right|_{i,j=1}^{2}\right]$ with
appropriate dimensions, define a Homographic transformation ${\rm\bf
H}_{m}(\Phi,\;P)$ as ${\rm\bf
H}_{m}(\Phi,\;P)=[\Phi_{11}P+\Phi_{12}][\Phi_{21}P+\Phi_{22}]^{-1}$.
Here, the matrix $\Phi_{21}P+\Phi_{22}$ is assumed to be square and
of full rank. This matrix transformation has been proved very useful
in solving many theoretical problems in systems and control, such as
the $H_{\infty}$ control problem, convergence analysis of Riccati
recursions, etc. \cite{kimura97,Bougerol93,ksh00}. An attractive
property of this transformation lies in its simplicity in
representing cascade connections, which is given in the following
lemma and can be obtained through straightforward algebraic
manipulations. This property plays important roles in analyzing the
asymptotic properties of the PCM $P_{t|t}$.

\hspace*{-0.4cm}{\bf Lemma 1.}\cite{kimura97} Assume that matrices
$\Phi_{1}$, $\Phi_{2}$ and $P$ have compatible dimensions. Moreover,
assume that all the required matrix inverses exist. Then, ${\rm\bf
H}_{m}(\Phi_{2},\;{\rm\bf H}_{m}(\Phi_{1},\;P))={\rm\bf
H}_{m}(\Phi_{2}\Phi_{1},\;P)$.

On the other hand, a matrix $\Phi=[\Phi_{ij}|_{i,j=1}^{2}]$ with
$\Phi_{ij}\in{\cal R}^{n\times n}$, $i,j=1,2$, is called Hamiltonian
if it satisfies $\Phi^{T}J\Phi=J$, in which $J=[{\rm\bf
col}\{0,\;-I_{n}\}, {\rm\bf col}\{I_{n},\;0\}]$. Hamiltonian
matrices are well encountered in optimal estimation and control, and
their characteristics have been extensively studied
\cite{Bougerol93,ksh00,kimura97}. Moreover, define four subsets of
Hamiltonian matrices ${\cal H}$, ${\cal H}_{l}$, ${\cal H}_{r}$ and
${\cal H}_{lr}$ respectively as ${\cal
H}=\left\{\;\Phi\;\left|\;\Phi=\left[\Phi_{ij}\right]_{i,j=1}^{2},\;
\Phi_{ij}\in{\cal R}^{n\times n},\;\Phi^{T}J\Phi=J,\;\Phi_{11}\;{\rm
invertible},\; \Phi_{12}\Phi_{11}^{T}\geq
0,\;\Phi_{11}^{T}\Phi_{21}\geq 0 \;\right.\right\}$, ${\cal
H}_{lr}=\left\{\;\Phi\;\left|\;\Phi\in{\cal H},\;
\Phi_{12}\Phi_{11}^{T}>0,\;\Phi_{11}^{T}\Phi_{21}> 0
\;\right.\right\}$, ${\cal
H}_{l}=\left\{\;\Phi\;\left|\;\Phi\in{\cal
H},\;\Phi_{11}^{T}\Phi_{21}> 0 \right.\;\right\}$ and ${\cal
H}_{r}=\left\{\;\Phi\;\left|\;\right.\right.$
$\left.\left.\Phi\in{\cal H},\; \Phi_{12}\Phi_{11}^{T}>
0\;\right.\right\}$. Then, from their definitions, it can be
straightforwardly declared that ${\cal H}_{l}\subset {\cal H}$,
${\cal H}_{r}\subset {\cal H}$, ${\cal H}_{lr}\subset {\cal H}$ and
${\cal H}_{lr}={\cal H}_{r}\cap{\cal H}_{l}$.

The following properties of Hamiltonian matrices are given in
\cite{Bougerol93}, which are repeatedly used in the remaining
theoretical studies of this paper.

\hspace*{-0.4cm}{\bf Lemma 2.}\cite{Bougerol93} Assume that all the
involved matrices have compatible dimensions. Then, among elements
of the sets ${\cal H}$, ${\cal H}_{l}$, ${\cal H}_{r}$ and ${\cal
H}_{lr}$, and (semi-)PDMs, the following relations exist.
\begin{itemize}

\item if $\Phi_{1}\in{\cal H}$ and $\Phi_{2}\in{\cal H}$ (or ${\cal H}_{l}$, or ${\cal H}_{r}$, or ${\cal H}_{lr}$), then, both $\Phi_{2}\Phi_{1}$ and $\Phi_{1}\Phi_{2}$ belongs to ${\cal H}$ (or ${\cal H}_{l}$, or ${\cal H}_{r}$, or ${\cal H}_{lr}$);

\item Assume that $\Phi_{i}=\left[\left.\Phi_{i,pq}\right|_{p,q=1}^{2}\right]\in{\cal H}$, $i=1,2,\cdots,m$. Then,
\begin{itemize}
\item $\prod_{i=m}^{1}\Phi_{i}\in{\cal H}_{l}$ if and
only if
\begin{equation}
{\rm\bf
D}_{et}\left\{\Phi_{1,11}^{T}\Phi_{1,21}+\sum_{i=2}^{m}\left[\left(\prod_{k=1}^{i}\Phi_{k,11}^{T}\right)\Phi_{i,21}\left(\prod_{k=i-1}^{1}\Phi_{k,11}\right)\right]\right\}\neq
0
\end{equation}

\item $\prod_{i=m}^{1}\Phi_{i}\in{\cal H}_{r}$ if and only if
\begin{equation}
{\rm\bf
D}_{et}\left\{\sum_{i=1}^{m-1}\left[\left(\prod_{k=m}^{i+1}\Phi_{k,11}\right)\Phi_{i,12}\left(\prod_{k=i}^{m}\Phi_{k,11}^{T}\right)\right]+\Phi_{m,12}\Phi_{m,11}^{T}\right\}\neq
0
\end{equation}
\end{itemize}

\item Assume that $\Phi\in{\cal H}$. Then, for an arbitrary $P\geq 0$, ${\rm\bf H}_{m}(\Phi,\;P)$ is well defined
and is at least a semi-PDM. If in addition that ${\rm\bf
D}_{et}(P)\neq 0$, then ${\rm\bf D}_{et}\left\{{\rm\bf
H}_{m}(\Phi,\;P)\right\}$ is also positive;

\item Assume that $\Phi\in{\cal H}_{lr}$. Then, for every $P\geq 0$, ${\rm\bf H}_{m}(\Phi,\;P)$ is certainly a PDM;

\item Assume that $\Phi\in{\cal H}$. Then, $\delta\left\{{\rm\bf
H}_{m}(\Phi,\;P),\;{\rm\bf H}_{m}(\Phi,\;Q)\right\}\leq\delta(P,Q)$,
whenever $P,\;Q>0$;

\item Assume that $\Phi\in{\cal H}_{l}$ or $\Phi\in{\cal H}_{r}$. Then, for any $P,\;Q>0$, $\delta\left\{{\rm\bf
H}_{m}(\Phi,\;P),\;\right.$ $\left.{\rm\bf
H}_{m}(\Phi,\;Q)\right\}<\delta(P,Q)$;

\item Assume that $\Phi\in{\cal H}_{lr}$. Then, there exists a $\rho(\Phi)$ belonging to $(0,\;1)$, such that for all $P,\;Q>0$, $\delta\left\{{\rm\bf
H}_{m}(\Phi,\;P),\;{\rm\bf
H}_{m}(\Phi,\;Q)\right\}\leq\rho(\Phi)\delta(P,Q)$.
\end{itemize}

To analyze asymptotic properties of RSEIO, the following results on
iterated functions governed by a semi-Markov process are also
needed, which have been successfully applied to establishing
convergence properties of KFIO \cite{Censi11,be88,stenflo96}.

\hspace*{-0.4cm}{\bf Lemma 3.}\cite{stenflo96} Let $f_{i}(\cdot)$,
$i=1,2,\cdots,p$, be a map from a metric space $({\cal X},\;\rho)$
to itself, and $I_{k}|_{k=1}^{\infty}$ a semi-Markov chain taking
values only from the set $\{\;1,\;2,\;\cdots,\; p\;\}$. Denote the
renewal process related to $I_{k}|_{k=1}^{\infty}$ by
$(s_{i},\;\delta_{i})|_{i=1}^{\infty}$, and the departure of $k$
from the last renewal by $t_{k}$. Assume that
$s_{i}|_{i=1}^{\infty}$ is irreducible,
$(I_{k},\;t_{k})|_{k=1}^{\infty}$ is aperiodic, and ${\rm\bf
E}(\delta_{i})<\infty$. If there exists an integer $N\geq 1$, such
that
\begin{equation}
{\rm\bf E}_{\{I_{i}|_{i=1}^{N}\}}\left\{{\rm
log\:L}_{ip}\left[f_{I_{1}}(f_{I_{2}}(\:\cdots\:f_{I_{N}}(\cdot)\:\cdots))\right]\right\}<0
\end{equation}
Then, the recursive random walk $(I_{k},\; X_{k})|_{k=1}^{\infty}$
with $X_{k}=f_{I_{k}}(X_{k-1})$ has a unique stationary
distribution. Moreover, for any initial $(I_{0},\;X_{0})$, the
empirical distribution tends to this stationary distribution with
probability one.

\subsection{Convergence Analysis}

To utilize the results of the previous subsection, $P_{t+1|t+1}$
should be expressed as a Homographic transformation of $P_{t|t}$.
When no information is contained in $y_{t+1}$ about the plant
output, RSEIO performs a Lyapunov recursion using nominal system
matrices, which makes it straightforward to establish this expected
relation. However, when the received signal $y_{t+1}$ contains
information about plant output, although the estimation is still
similar to that of the Kalman filter, the relation between
$P_{t+1|t+1}$ and $P_{t|t}$ is quite complicated. This means that to
clarify the asymptotic characteristics of the PCM $P_{t|t}$, another
recursive form is required for it under the situation
$\gamma_{t+1}=1$.

Note that in the convergence analysis of the sensitivity
penalization based RSE, a relatively compact relation between
$P_{t+1|t}$ and $P_{t|t-1}$ has been established in \cite{zhou11}.
However, this relation is not very convenient in deriving the
required relation between $P_{t+1|t+1}$ and $P_{t|t}$. In this
paper, we take a different approach in establishing this relation,
which is given in the next theorem and whose proof is deferred to
the appendix.

\hspace*{-0.4cm}{\bf Theorem 2.} Denote the matrix
$A_{t}(0)-\lambda_{t}B_{t}(0)(Q_{t}^{-1}+\lambda_{t}T_{t}^{T}T_{t})^{-1}T_{t}^{T}S_{t}$
by $\check{A}_{t}$ and assume it is invertible. Define matrices
$\tilde{A}_{t}$, $\tilde{B}_{t}$, $\tilde{C}_{t+1}$, $\tilde{Q}_{t}$
and $\tilde{R}_{t+1}$ respectively as follows,
\begin{eqnarray*}
& &
\tilde{A}_{t}=\check{A}_{t}+B_{t}(0)\check{Q}_{t}\tilde{B}_{t}^{T}\tilde{S}_{t}^{T}\tilde{S}_{t},\hspace{0.5cm}
\tilde{B}_{t}=\check{A}_{t}^{-1}B_{t}(0),\hspace{0.5cm}\tilde{Q}_{t}=\check{Q}_{t}
+\check{Q}_{t}\tilde{B}_{t}^{T}\tilde{S}_{t}^{T}\tilde{S}_{t}\tilde{B}_{t}\check{Q}_{t}\\
& &
\tilde{S}_{t}\!=\!\sqrt{\lambda_{t}}\left[I+\lambda_{t}T_{t}Q_{t}T_{t}^{T}\right]^{-1/2}S_{t},\hspace{0.25cm}
\tilde{C}_{t+1}\!=\!\left[\begin{array}{c}
\tilde{S}_{t}\check{A}_{t}^{-1} \\ C_{t+1}(0)
\end{array}\right],\hspace{0.25cm}
\tilde{R}_{t+1}\!=\!\left[\begin{array}{cc}
I+\tilde{S}_{t}\tilde{B}_{t}\check{Q}_{t}\tilde{B}_{t}\tilde{S}_{t}^{T}
& 0 \\ 0 & R_{t+1} \end{array}\right]
\end{eqnarray*}
in which
$\check{Q}_{t}=(Q_{t}^{-1}+\lambda_{t}T_{t}^{T}T_{t})^{-1}$. If
$\gamma_{t+1}\neq 0$, then,
\begin{equation}
P_{t+1|t+1}^{-1}=\left[\tilde{A}_{t}P_{t|t}\tilde{A}_{t}^{T}+B_{t}(0)\tilde{Q}_{t}B_{t}^{T}(0)\right]^{-1}+\tilde{C}_{t+1}^{T}\tilde{R}^{-1}_{t+1}\tilde{C}_{t+1}
\end{equation}

Note that although the matrices $\tilde{A}_{t}$, $\tilde{C}_{t+1}$,
$\tilde{Q}_{t}$ and $\tilde{R}_{t+1}$ have a complicated form, all
of them are independent of system input-output data, and can
therefore be computed off-line. This also means
that the recursion formula for $P_{t+1|t+1}$ in Theorem 1 is more
suitable for performing robust state estimations, while that in
Theorem 2 matches better for its asymptotic property analysis. It is
also worthwhile to point out that invertibility of the matrix
$\check{A}_{t}$ is not required in deriving the RSE of Theorem 1,
which implies that further efforts are still required to establish
its asymptotic properties in the most general situation.

From Theorems 1 and 2, it is clear that depending on whether or not
$y_{t+1}$ contains information about plant outputs, the PCM
$P_{t+1|t+1}$ performs alternatively a Lyapunov recursion and a
Riccati recursion. This is very similar to that of KFIO. But as
robustness has been taken into account, system matrices in the
Riccati recursion are different from those in the Lyapunov
recursion. This difference significantly complicates convergence
analysis for RSEIO and makes its conclusions different from those of
KFIO.

Lyapunov and Riccati equations/recursions play
important roles in system analysis and synthesis, and their
properties have been extensively studied \cite{afij03,ksh00}. When
plant measurements are missed randomly, the alternative
Lyapunov/Riccati recursion in both the KFIO and the RESIO becomes a
random process, which makes its convergence analysis much more
mathematically difficult and some basic conclusions different from
their counterparts of deterministic recursions
\cite{Censi11,ms12,sem10,km12,ssfpjs04}. For example, in
\cite{ssfpjs04}, it is proved that for an unstable system,
simultaneous controllability and observability are no longer
sufficient for guaranteeing the boundedness of the covariance matrix
of estimation errors of the KFIO. It can also be seen in the
following analysis that when plant output measurement receiving
probability is greater than $0$, controllability and observability
are {\it only} a sufficient condition for the convergence of the
RESIO. On the other hand, as system matrices in the Riccati
recursion are different from those in the Lyapunov recursion in
RESIO, its convergence analysis is more mathematically involved than
that of KFIO.

To simplify mathematical expressions in the following discussions,
$A_{t}(0)$ and $B_{t}(0)$ are respectively abbreviated to $A_{t}$
and $B_{t}$. Moreover, assume that both the matrix $A_{t}$ and the
matrix $\tilde{A}_{t}$ are invertible. Define matrix $\Phi_{t+1}$ as
\begin{equation}
\Phi_{t+1}=\left\{\begin{array}{ll} \left[\begin{array}{cc} A_{t} &
B_{t}Q_{t}B_{t}^{T}A_{t}^{-T} \\ 0 & A_{t}^{-T} \end{array}\right] &
\gamma_{t+1}=0 \\
\left[\begin{array}{cc} \tilde{A}_{t} & B_{t}\tilde{Q}_{t}B_{t}^{T}\tilde{A}_{t}^{-T} \\
\tilde{C}_{t+1}^{T}\tilde{R}_{t+1}^{-1}\tilde{C}_{t+1}\tilde{A}_{t}
&
[I+\tilde{C}_{t+1}^{T}\tilde{R}_{t+1}^{-1}\tilde{C}_{t+1}B_{t}\tilde{Q}_{t}B_{t}^{T}]\tilde{A}_{t}^{-T}
\end{array}\right] & \gamma_{t+1}=1 \end{array}\right.
\label{eqn:2}
\end{equation}
Then, straightforward algebraic manipulations show that $\Phi_{t+1}$
is always a Hamiltonian matrix, and always belongs to the set ${\cal
H}$. Moreover, the following results can be immediately obtained
from Lemmas 1 and 2, as well as Theorems 1 and 2.

\hspace*{-0.4cm}{\bf Corollary 1.} Assume that RSEIO starts from
$t=0$ with $\hat{x}_{0|0}$ and $P_{0|0}$. Moreover, assume that both
the matrix $A_{t}$ and the matrix $\tilde{A}_{t}$ are of full rank
at all the sampled time instants. Then, for an arbitrary semi-PDM
$P_{0|0}$ and an arbitrary time instant $t=1,2,\cdots$,
\begin{equation}
P_{t|t}={\rm\bf H}_{m}\left(\prod_{k=t}^{1}\Phi_{k},\;P_{0|0}\right)
\label{eqn:9}
\end{equation}

\hspace*{-0.4cm}{\bf Proof:} Note that $\Phi_{k}\in{\cal H}$,
$k=1,2,\cdots,t$. It can be declared from Lemma 2 that when both the
matrix $A_{k}$ and the matrix $\tilde{A}_{k}$ are invertible, the
Homographic transformation ${\rm\bf H}_{m}\left(\Phi_{k},\;P\right)$
is always well defined for every $n\times n$ dimensional semi-PDM
$P$.

From the definition of the matrix $\Phi_{k}$ and Theorems 1 and 2,
it is obvious that for every $k=0,1,\cdots,t-1$, no matter
$\gamma_{k+1}=0$ or $\gamma_{k+1}=1$, we always have that
\begin{equation}
P_{k+1|k+1}={\rm\bf H}_{m}\left(\Phi_{k+1},\;P_{k|k}\right)
\end{equation}
Hence, it can be claimed from Lemma 2 that when $P_{0|0}$ is a
semi-PDM, all the involved $P_{k|k}$s are well defined and are at
least a semi-PDM. Moreover, a repetitive utilization of Lemma 1
leads to,
\begin{eqnarray}
P_{t|t}&=&{\rm\bf H}_{m}\left(\Phi_{t},\;{\rm\bf
H}_{m}\left(\Phi_{t-1},\;\cdots,\;{\rm\bf
H}_{m}\left(\Phi_{1},\;P_{0|0}\right)\cdots\right)\right)\nonumber\\
&=&{\rm\bf H}_{m}\left(\Phi_{t}\Phi_{t-1},\;{\rm\bf
H}_{m}\left(\Phi_{t-2},\;\cdots,\;{\rm\bf
H}_{m}\left(\Phi_{1},\;P_{0|0}\right)\cdots\right)\right)\nonumber\\
&=& \cdots \nonumber\\
&=&{\rm\bf H}_{m}\left(\prod_{k=t}^{1}\Phi_{k},\;P_{0|0}\right)
\end{eqnarray}

This completes the proof. \hspace{\fill}$\Diamond$

Similar to the proof of Corollary 1, it can also be proved that for
every semi-PDM $X$, ${\rm\bf H}_{m}(\Phi_{1},\;{\rm\bf
H}_{m}(\Phi_{2},$ $\cdots,\;{\rm\bf
H}_{m}(\Phi_{t},\;X)\cdots))={\rm\bf
H}_{m}(\prod_{k=1}^{t}\Phi_{k},\;X)$.

In the rest of this paper, in order to explicitly express the
dependence of the matrix $\Phi_{t}$ on a realization of
$\gamma_{t}$, this matrix is sometimes, with a little abuse of
symbols, written as $\Phi_{R(t)}$ when necessary, in which
$R(t)|_{t=1}^{\infty}$ is a realization of the random process
$\gamma_{t}|_{t=1}^{\infty}$.

Having these preparations, we are ready to analyze asymptotic
properties of the PCM $P_{t|t}$. To perform this analysis, it is
assumed in the remaining of this section that the nominal model of
the plant, as well as the first order derivatives of the innovation
process $e_{t}(\varepsilon_{t},\varepsilon_{t+1})$ with respect to
every parametric modelling error, do not change with the variable
$t$. That is, $A_{t}(0)$, $B_{t}(0)$, $C_{t}(0)$, $R_{t}$, $Q_{t}$,
$S_{t}$ and $T_{t}$ are no longer a function of the temporal
variable $t$. Under this assumption, it is feasible to define
matrices $A^{[1]}$, $A^{[2]}$, $G^{[1]}$, $G^{[2]}$ and $H^{[1]}$,
all of which do not depend on the variable $t$, respectively as
\begin{displaymath}
A^{[1]}=\tilde{A}_{t},
\hspace{0.5cm}G^{[1]}=B_{t}{\tilde{Q}_{t}^{1/2}},\hspace{0.5cm}
H^{[1]}=\tilde{R}_{t+1}^{-1/2}\tilde{C}_{t+1},
 \hspace{0.5cm} A^{[2]}=A_{t},
\hspace{0.5cm} G^{[2]}=B_{t}Q_{t}^{1/2}
\end{displaymath}

Using these symbols, it can be straightforwardly proved that
$B_{t}Q_{t}B_{t}^{T}=G^{[2]}G^{[2]T}$,
$B_{t}\tilde{Q}_{t}B_{t}^{T}=G^{[1]}G^{[1]T}$ and
$\tilde{C}_{t+1}^{T}\tilde{R}_{t+1}^{-1}\tilde{C}_{t+1}=H^{[1]T}H^{[1]}$.
On the basis of these relations and Lemmas 1 and 2, the following
conclusions are obtained on the product of matrices
$\Phi_{i}|_{i=1}^{N}$ for an arbitrary positive integer $N$. Their
proof is given in the appendix.

\hspace*{-0.4cm}{\bf Theorem 3.} For a prescribed positive integer
$N$, $\prod_{t=1}^{N}\Phi_{t}\in {\cal H}_{l}$ if and only if there
exists an integer sequence $t_{i}|_{i=0}^{p}$ satisfying
$0=t_{0}<1\leq t_{1}<t_{2}<\cdots<t_{p}\leq N$, such that the matrix
$O_{b}$ is of full column rank which is defined as
\begin{equation}
O_{b}={\rm\bf col}\left\{H^{[1]},\;
H^{[1]}A^{[1]}(A^{[2]})^{t_{p}-t_{p-1}-1},\; \cdots,\;
H^{[1]}\prod_{j=1}^{p}A^{[1]}(A^{[2]})^{t_{j}-t_{j-1}-1}\right\}
\label{eqn:11}
\end{equation}

When the subset ${\cal H}_{r}$ is concerned, we have the following
results. Their proof is also deferred to the appendix.

\hspace*{-0.4cm}{\bf Theorem 4.} For a prescribed positive integer
$N$, $\prod_{t=1}^{N}\Phi_{t}\in {\cal H}_{r}$ if and only if there
exists an integer sequence $t_{i}|_{i=0}^{p+1}$ satisfying
$0=t_{0}<1\leq t_{1}<t_{2}<\cdots<t_{p}<t_{p+1}=N+1$, such that the
matrix $C_{n}$ defined as
\begin{equation}
C_{n}=\left[(A^{[2]})^{t_{1}-1}C_{n,0}\;\;
C_{n,1}\;\;(A^{[2]})^{t_{1}-1}A^{[1]}\left[C_{n,2}\;\;\cdots\;\;
\left(\prod_{s=1}^{p-1}(A^{[2]})^{t_{s+1}-t_{s}-1}A^{[1]}\right)C_{n,p+1}
\right]\right]
\label{eqn:12}
\end{equation}
is of full row rank, in which
\begin{eqnarray*}
& &
C_{n,0}=\left[G^{[1]}\;\;A^{[1]}(A^{[2]})^{t_{2}-t_{1}-1}G^{[1]}\;\;\cdots\;\;
\left(\prod_{s=1}^{p-1}A^{[1]}(A^{[2]})^{t_{s+1}-t_{s}-1}\right)G^{[1]}\right] \\
& & C_{n,i}=\left[G^{[2]}\;\;A^{[2]}G^{[2]}\;\; \cdots\;\;
(A^{[2]})^{t_{i}-t_{i-1}-2}G^{[2]}\right],\hspace{0.5cm}i=1,2,\cdots,p+1
\end{eqnarray*}

From these results, some sufficient conditions can be obtained for
the existence of a finite positive integer $N$, such that a map
defined in a similar way as that of  Equation (\ref{eqn:9}) is
strictly contractive.

\hspace*{-0.4cm}{\bf Corollary 2.} There exists a finite binary
sequence $R^{[N]}(t)|_{t=1}^{N}$ with $N$ a finite positive integer,
such that the corresponding matrices $\Phi_{R^{[N]}(t)}|_{t=1}^{N}$
satisfy
\begin{itemize}
\item $\prod_{t=1}^{N}\Phi_{R^{[N]}(t)}$ belongs to ${\cal H}_{l}$, if there
exists an integer $m$ belonging to $[0,\;n-1]$, such that the matrix
pair $(A^{[1]}(A^{[2]})^{m},\;H^{[1]})$ is observable.
\item $\prod_{t=1}^{N}\Phi_{R^{[N]}(t)}$ belongs to ${\cal H}_{r}$, if one of
the following conditions are satisfied.
\begin{itemize}
\item there
exists an integer $m$ belonging to $[0,\;n-1]$, such that the matrix
pair $(A^{[1]}(A^{[2]})^{m},\;G^{[1]})$ is controllable;
\item the
matrix pair $(A^{[2]},\;G^{[2]})$ is controllable;
\item there
exists an integer $m$ belonging to $[0,\;n-1]$, such that the matrix
pair $((A^{[2]})^{m}A^{[1]},\;G^{[2]})$ is controllable.
\end{itemize}
\item $\prod_{t=1}^{N}\Phi_{R^{[N]}(t)}$ belongs to ${\cal H}_{lr}$, if both
the above observability condition and one of the above
controllability conditions are satisfied simultaneously.
\end{itemize}

\hspace*{-0.4cm}{\bf Proof:} Assume that there exists an integer
$m$, such that $0\leq m\leq n-1$ and the matrix pair
$(A^{[1]}(A^{[2]})^{m},\;H^{[1]})$ is observable. Designate $N$ and
$t_{i}$ respectively as  $N=(n-1)(m-1)+1$ and $t_{i}=(i-1)*(m+1)+1$,
$1\leq i\leq n-1$. Then, $N$ is of a finite value. Moreover, from
the observability of $(A^{[1]}(A^{[2]})^{m},\;H^{[1]})$ and the
definition of the matrix $O_{b}$ in Equation (\ref{eqn:11}), it can
be declared that the matrix $O_{b}$ is of full column rank. It can
therefore be claimed from Theorem 3 that
$\prod_{t=1}^{N}\Phi_{R^{[N]}(t)}\in {\cal H}_{l}$.

Note that both the matrix $A^{[1]}$ and the matrix $A^{[2]}$ are
assumed invertible. It can therefore be declared from the definition
of the matrix $C_{n}$ in Equation (\ref{eqn:12}) that, $C_{n}$ is of
full row rank if any of the matrices $C_{n,i}$, $i=0,1,\cdots,p+1$,
has this property. The remaining arguments are similar to those for
showing the existence of a finite integer $N$ such that
$\prod_{t=1}^{N}\Phi_{R^{[N]}(t)}\in {\cal H}_{l}$, and are
therefore omitted.

From the definitions of the sets ${\cal H}_{l}$, ${\cal H}_{r}$ and
${\cal H}_{lr}$, it is obvious that a matrix $\Phi$ belongs to
${\cal H}_{lr}$ if and only if it simultaneously belongs to both
${\cal H}_{l}$ and ${\cal H}_{r}$. On the other hand, if there exist
positive integers $N_{*}$ and $N$ with $N_{*}<N$ such that
$\prod_{t=1}^{N_{*}}\Phi_{R^{[N]}(t)}\in {\cal H}_{l}$ and
$\prod_{t=N_{*}+1}^{N}\Phi_{R^{[N]}(t)}\in {\cal H}_{r}$, then, it
can be claimed from Lemma 2 that $\prod_{t=1}^{N}\Phi_{R^{[N]}(t)}$
belongs to both the set ${\cal H}_{l}$ and the set ${\cal H}_{r}$,
and therefore $\prod_{t=1}^{N}\Phi_{R^{[N]}(t)}\in {\cal H}_{lr}$.
Similarly, if there exist positive integers $N_{*}$ and $N$ with
$N_{*}<N$ such that $\prod_{t=1}^{N_{*}}\Phi_{R^{[N]}(t)}\in {\cal
H}_{r}$ and $\prod_{t=N_{*}+1}^{N}\Phi_{R^{[N]}(t)}\in {\cal
H}_{l}$, then, $\prod_{t=1}^{N}\Phi_{R^{[N]}(t)}$ also belongs to
the set ${\cal H}_{lr}$. The conclusions about the existence of a
finite integer $N$ such that $\prod_{t=1}^{N}\Phi_{R^{[N]}(t)}\in
{\cal H}_{lr}$ are therefore straightforward results of those for
$\prod_{t=1}^{N}\Phi_{R^{[N]}(t)}\in {\cal H}_{l}$ and
$\prod_{t=1}^{N}\Phi_{R^{[N]}(t)}\in {\cal H}_{r}$.

This completes the proof. \hspace{\fill}$\Diamond$

In the above proof, a periodic $R^{[N]}(t)|_{t=1}^{N}$ is
constructed to derive conditions for the existence of a finite
integer $N$ such that $\prod_{t=1}^{N}\Phi_{t}$ belongs respectively
to the sets ${\cal H}_{l}$, ${\cal H}_{r}$ and ${\cal H}_{lr}$.
These conditions are generally conservative but are simple to
verify, noting that both controllability and observability are
wildly accepted concepts in system analysis and synthesis, and
various efficient methods have been developed to check these
properties for a given dynamic system. If the matrices $A^{[1]}$ and
$A^{[2]}$ have the property that $A^{[1]}A^{[2]}=A^{[2]}A^{[1]}$,
then, less conservative results can be derived. The details are
omitted due to space considerations. These conditions are very
important in investigating asymptotic properties of RSEIO, which
becomes clear in the following Theorem 5. It remains interesting to
establish less conservative but easily verifiable conditions for the
existence of a finite integer $N$, such that the matrices $O_{b}$ in
Equation (\ref{eqn:11}) and $C_{n}$ in Equation (\ref{eqn:12}) are
respectively of full column rank and of full row rank.

On the other hand, if $A^{[1]}=A^{[2]}$ and $G^{[1]}=G^{[2]}$ are
simultaneously satisfied, then, it is straightforward to show that
the matrix $O_{b}$ in Equation (\ref{eqn:11}) is of full column rank
if and only if the matrix pair $(A^{[1]},\;H^{[1]})$ is observable,
while the matrix $C_{n}$ in Equation (\ref{eqn:12}) is of full row
rank if and only if the matrix pair $(A^{[1]},\;G^{[1]})$ is
controllable. This means that if the dynamic system $\rm\bf\Sigma$
is time invariant and its state space model is accurate, then, the
existence of a finite positive integer $N$ such that the matrix
product $\prod_{t=1}^{N}\Phi_{t}$ belongs to the set ${\cal H}_{lr}$
is equivalent to its simultaneous controllability and observability,
which is consistent with that reported in \cite{Bougerol93,Censi11}.

To investigate the asymptotic property of RSEIO, probability should
be investigated about the existence of strictly contractive mappings
among the random MVFs defined in a similar way as that of Equation
(\ref{eqn:9}). For this purpose, some symbols are introduced which
are some modifications of those adopted in \cite{rmf11}. Let
$\Gamma^{[N]}$ represent a finite random sequence
$\gamma_{t}|_{t=1}^{N}$ with $\gamma_{t}$ takes values only from the
set $\{0,\;1\}$. Let ${\cal S}^{[N]}$ denote the set consisting of
all binary sequences of length $N$, that is,
\begin{displaymath}
{\cal
S}^{[N]}=\left\{\;S_{m}^{[N]}\;\left|\;S_{m}^{[N]}=\{\;S_{m}^{[N]}(i)|_{i=1}^{N}\;\},\;
S_{m}^{[N]}(i)\in\{0,\;1\},\;m=\sum_{i=1}^{N}2^{i-1}S_{m}^{[N]}(i)\;\right.\right\}
\end{displaymath}
Then, it is clear that the set ${\cal S}^{[N]}$ have exactly $2^{N}$
elements, and every element is a realization of the finite random
sequence $\Gamma^{[N]}$.

The following results are some extensions and modifications of those
of \cite{rmf11}. Their proof is given in the appendix.

\hspace*{-0.4cm}{\bf Lemma 4.} For an arbitrary positive integer
$N$, let $S_{m}^{[N]}$ denote the $m+1$-th element of the set ${\cal
S}^{[N]}$. Then,
\begin{itemize}
\item if the stochastic sequence $\gamma_{t}|_{t=1}^{\infty}$ is a
series of independent random variables with the Bernoulli
distribution of a constant expectation $\bar{\gamma}$, then,
\begin{equation}
{\rm log}\left[{\rm\bf
P}_{r}\left(\Gamma^{[N]}=S_{m}^{[N]}\right)\right]={\rm
log}(\bar{\gamma})\sum_{i=1}^{N}S_{m}^{[N]}(i)+{\rm
log}(1-\bar{\gamma})\left(N-\sum_{i=1}^{N}S_{m}^{[N]}(i)\right)
\end{equation}
\item if the random sequence $\gamma_{t}|_{t=1}^{\infty}$ is a
Markov chain with a transition probability matrix $[{\rm\bf
col}\{\alpha,\;  1-\alpha\},\; {\rm\bf col}\{1-\beta,\; \beta\}]$
and ${\rm\bf P}_{r}(\gamma_{0}=1)=\bar{\gamma}$, in which both
$\alpha$ and $\beta$ belong to $(0,1)$. Then,
\begin{eqnarray}
{\rm log}\!\left[{\rm\bf
P}_{r}\!\left(\!\Gamma^{[N]}\!=\!S_{m}^{[N]}\right)\!\right]\!\!\!\!&=&\!\!\!\!(N\!-\!1){\rm
log}(\beta)\!+\!{\rm
log}\!\left(\!\frac{1\!-\!\alpha}{\beta}\!\right)\!\sum_{k=1}^{N-1}\!S_{m}^{[N]}(k)\!+\!{\rm
log}\!\left(\!\frac{1}{\beta}\!-\!1\!\right)\!\sum_{k=2}^{N}S_{m}^{[N]}(k)\!+\nonumber\\
& &\!\!\!\!\hspace*{2cm}{\rm
log}\!\left(\!\frac{\alpha\beta}{(1\!-\!\alpha)(1\!-\!\beta)}\!\right)\!\sum_{k=2}^{N}\!\left[\!S_{m}^{[N]}(k)S_{m}^{[N]}(k\!-\!1)\!\right]\!+\nonumber\\
& &\!\!\!\! \hspace*{2cm}{\rm
log}\!\left\{\!S_{m}^{[N]}(1)\!+\![1\!-\!2S_{m}^{[N]}(1)][\beta\!+\!\bar{\gamma}(1\!-\!\alpha\!-\!\beta)]\!\right\}
\end{eqnarray}
\end{itemize}

Lemma 4 makes it clear that for an identically and independently
distributed (i.i.d.) Bernoulli process, if its expectation is
greater than $0$, then, for any positive integer $N$ and any element
$S_{m}^{[N]}$ of the set ${\cal S}^{[N]}$ that does not take a
constant value, the probability that the random sequence
$\Gamma^{[N]}$ has a realization $S_{m}^{[N]}$ is greater than $0$.
That is, when $\bar{\gamma}>0$, except the element $S_{m}^{[N]}$
with $m=0$ or $m=2^{N}-1$, every other element of the set ${\cal
S}^{[N]}$ has a positive probability to become a realization of the
random sequence $\Gamma^{[N]}$. On the other hand, when the random
sequence $\gamma_{t}$ is described by a Markov chain, then, if
$0<\alpha,\;\beta<1$, every element of the set ${\cal S}^{[N]}$ with
$m=1,2,\cdots,2^{N}-2$, can also be realized by the random sequence
$\Gamma^{[N]}$ with a positive probability.

Similar results can be derived for situations in which random
measurement droppings are described by other stochastic process,
such as a semi-Markov chain, etc. The details are not included for
space considerations.

From the above results, a convergence property can be established
for the PCM $P_{t|t}$ of RSEIO. Its proof is provided in the
appendix.

\hspace*{-0.4cm}{\bf Theorem 5.} For the dynamic system
${\rm\bf\Sigma}$ with $A_{t}(0)$, $\check{A}_{t}$ and
$\tilde{A}_{t}$ invertible, assume that there exist two positive
integers $m_{1}$ and $m_{2}$ such that the matrix pair
$(A^{[1]}(A^{[2]})^{m_{1}},\;H^{[1]})$ is observable and one of the
following three conditions is satisfied,
\begin{itemize}
\item the matrix pair
$(A^{[1]}(A^{[2]})^{m_{2}},\;G^{[1]})$ is controllable;
\item the matrix pair
$((A^{[2]})^{m_{2}}A^{[1]},\;G^{[2]})$ is controllable;
\item the matrix pair $(A^{[2]},\;G^{[2]})$ is
controllable.
\end{itemize}
Then, the PCM $P_{t|t}$ of RSEIO converges to a stationary
distribution with probability one that is independent of its initial
value $P_{0|0}$, provided that one of the following two conditions
is satisfied by the random measurement dropping process
$\gamma_{t}$,
\begin{itemize}
\item At every sampled time instant $t$, the random dropping
is an i.i.d. Bernoulli variable with a positive expectation;
\item The random dropping process can be described by a Markov chain
with a transition probability matrix $[{\rm\bf col}\{\alpha,\;
1-\alpha\},\; {\rm\bf col}\{1-\beta,\; \beta\}]$ and
$0<\alpha,\;\beta<1$.
\end{itemize}

The above theorem gives some sufficient conditions for the
convergence of the PCM $P_{t|t}$ of RSEIO. Note that for a $n\times
n$ dimensional matrix $A$, from the Hamiltonian-Cayley theorem
\cite{hj91}, we know that $A^{k}$ with any $k\geq n$ can be
expressed as a linear combination of $A^{i}$, $i=0,1,\cdots,n-1$.
From this result and the discussions after Corollary 2,
straightforward algebraic manipulations show that if the dynamic
system $\rm\bf\Sigma$ is time invariant and has an accurate state
space model, then, simultaneous observability of the matrix pair
$(A^{[1]},\;H^{[1]})$ and controllability of the matrix pair
$(A^{[1]},\;G^{[1]})$ are in fact necessary and sufficient condition
on the system matrices. These mean that the conditions of Theorem 5
reduce to those of \cite{Censi11,km12} in which asymptotic
properties of the covariance matrix is investigated for KFIO.
However, when there are modelling errors,
observability of $(A^{[1]},\;H^{[1]})$ and controllability of
$(A^{[1]},\;G^{[1]})$ or $(A^{[2]},\;G^{[2]})$ are {\it only}
sufficient conditions. This implies that more opportunities exist
for the convergence of the PCM $P_{t|t}$ when the plant system
matrices are not accurate.

Note that the gain matrix of RSEIO is equal to
$P_{t|t}C_{t}(0)R_{t}^{-1}$ at the time instant $t$ when $y_{t}$
contains information about plant output, and is equal to $0$ in
other situations. Sufficient conditions can be derived directly from
Theorem 5 for the convergence of this gain matrix. On the other
hand, it is worthwhile to point out that estimation accuracy is a very important performance index for estimators, which is usually reflected by the
covariance matrix of estimation errors. While the PCM of the RSEIO is closely related to the covariance matrix of its estimation errors, these two matrices are not equal to each other in general. It is expected that through some arguments similar to those of \cite{zhou11}, some asymptotic
properties can be established for an upper bound of the
covariance matrix of estimation errors of RSEIO. This establishment, of course, requires some assumptions on the parametric modelling errors, such as their variation intervals and/or
statistical distributions, etc. This is an interesting issue under current investigations. Due to space
considerations, detailed discussions are omitted.

Results of Theorem 5 can be easily extended to other descriptions of
the random measurement dropping process. However, this theorem only
establishes existence of a stationary distribution for the PCM
matrix $P_{t|t}$. Further efforts are still required to derive an
explicit expression for this stationary distribution.

\section{A Numerical Example}

To illustrate estimation performances of the developed estimation
algorithm, some numerical simulation results are reported in this
section. The plant is selected to be the same as that of \cite{lz11}
which has the following system matrices, initial conditions, and
covariance matrices for process noises and measurement errors,
respectively.
\begin{eqnarray*}
& & A_{t}(\varepsilon_{t})\!=\!\left[\!\!\begin{array}{cc} 0.9802 & 0.0196 \\
0 & 0.9802 \end{array}\!\!\right]\!+\!\left[\!\!\begin{array}{c} 0.0198 \\
0
\end{array}\!\!\right]\!\varepsilon_{t}\left[0\;\; 5\right],\hspace{0.2cm}
B_{t}(\varepsilon_{t})\!=\!\left[\!\!\begin{array}{cc} 1 & 0 \\ 0 &
1
\end{array}\!\!\right],\hspace{0.2cm} Q_{t}\!=\!\left[\!\!\begin{array}{cc}
1.9608 & 0.0195 \\ 0.0195 & 1.9605 \end{array}\!\!\right] \\
& & C_{t}(\varepsilon_{t})=[1\;\; -1],\hspace{0.5cm}
R_{t}=1,\hspace{0.5cm} {\rm\bf
E}\{x_{0}\}=[1\;\;0]^{T},\hspace{0.5cm} P_{0}=I_{2}
\end{eqnarray*}
in which $\varepsilon_{t}$ stands for a time varying parametric
error that is independent of each other and has a uniform
distribution over the interval $[-\delta,\;\delta]$. The measurement
dropping process $\gamma_{t}$ is assumed to be a stationary
Bernoulli process with its expectation equal to $0.8$. To compare
estimation accuracy of different methods, the estimator design
parameter $\mu_{t}$ is at first selected to be the same as that of
\cite{lz11}, that is, $\mu_{t}\equiv 0.8$.

\begin{figure}[!ht]
\begin{center}
\hspace*{-0.4cm}\includegraphics[width=2.2in]{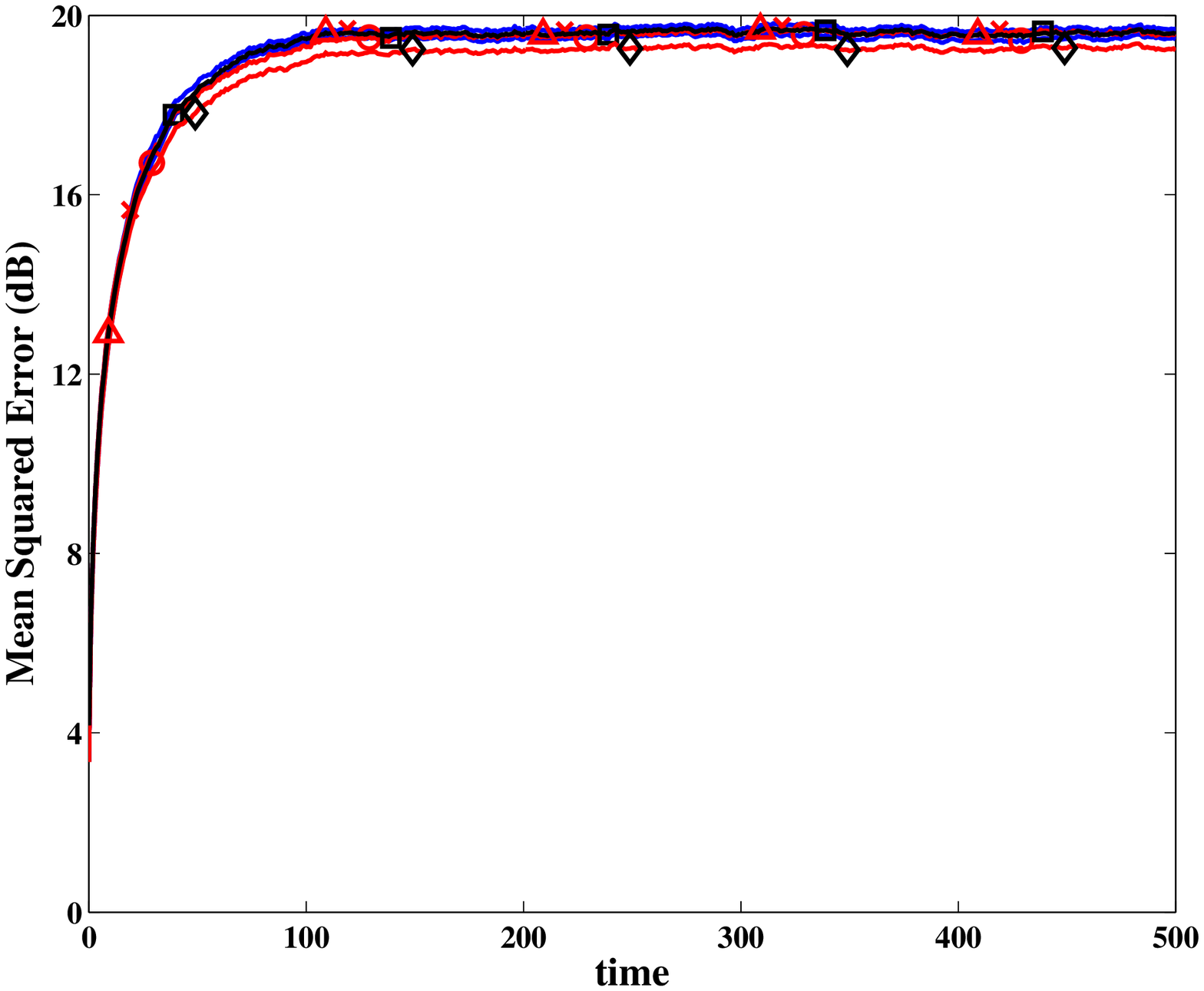}
\hspace{0.4cm}\includegraphics[width=2.2in]{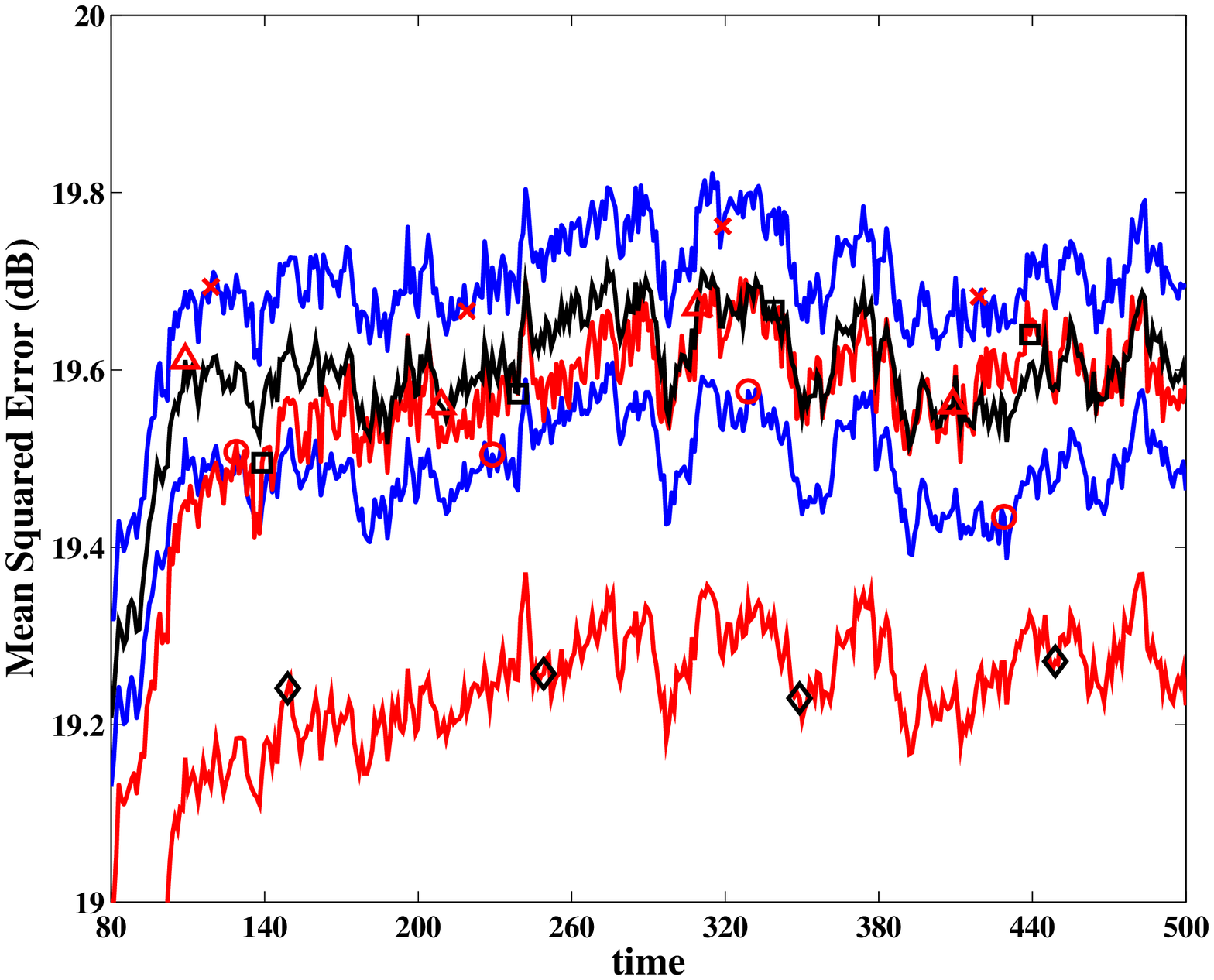}

\vspace{-0.2cm} \hspace*{-0.5cm} {\footnotesize (a)
$t\in[0,\;500],\;\delta=1,\;\mu_{t}\equiv 0.8$} \hspace{1.0cm}
{\footnotesize (b) $t\in[80,\;500],\;\delta=1,\;\mu_{t}\equiv 0.8$}

\hspace*{-0.4cm}\includegraphics[width=2.2in]{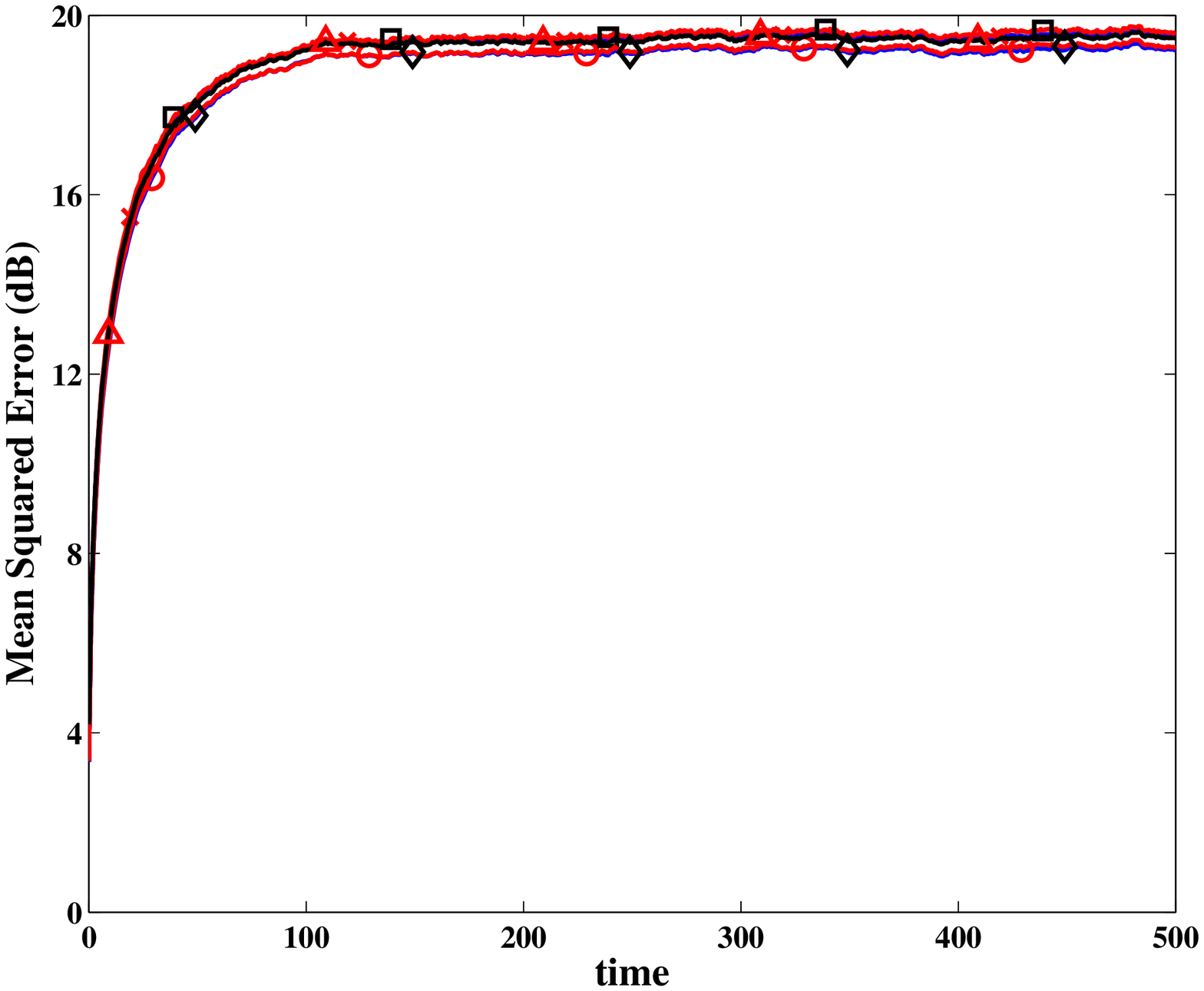}
\hspace{0.4cm}\includegraphics[width=2.2in]{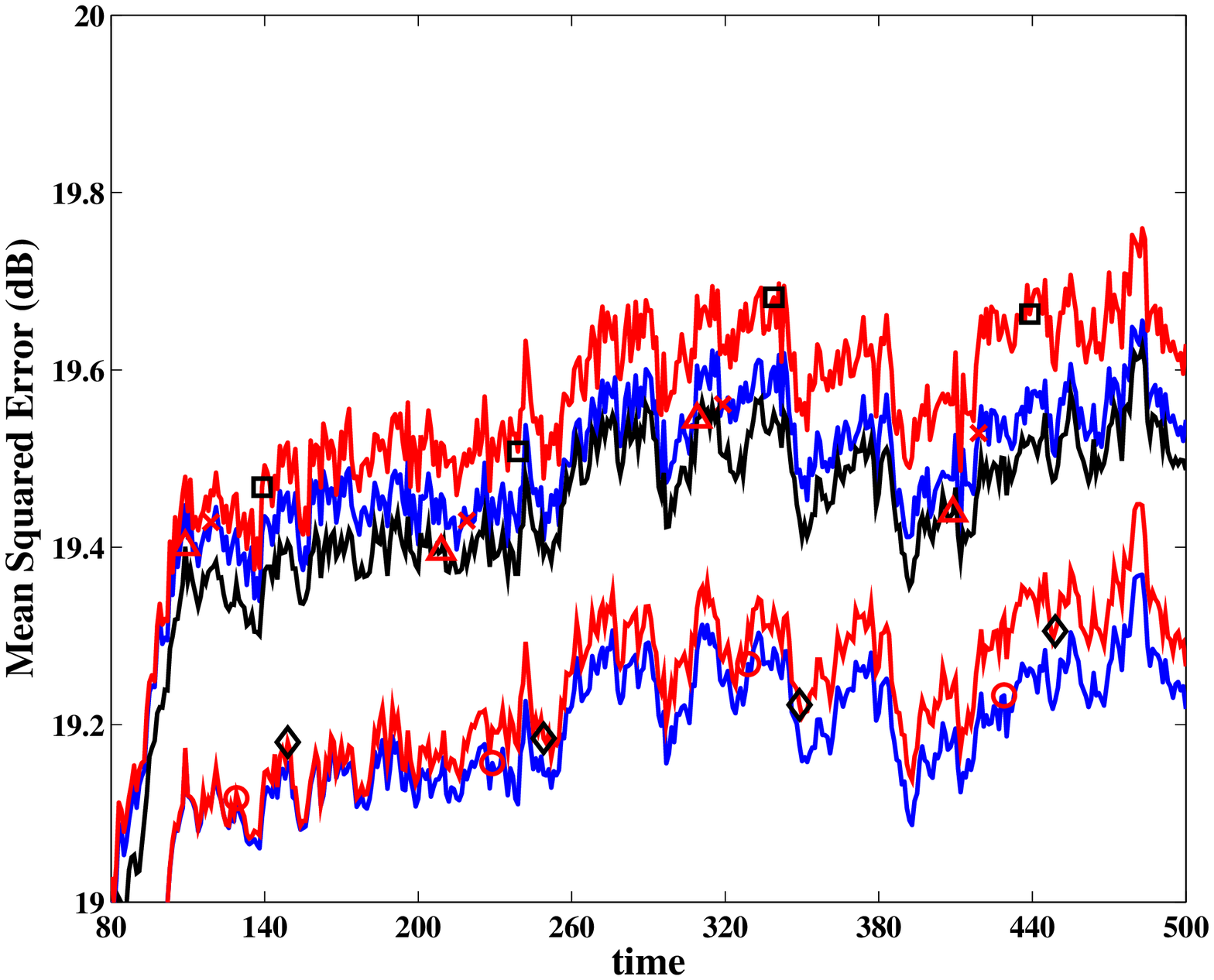}

\vspace{-0.2cm} \hspace*{-0.5cm} {\footnotesize (c)
$t\in[0,\;500],\;\delta=1,\;\mu_{t}\equiv 0.95$} \hspace{1.0cm}
{\footnotesize (d) $t\in[80,\;500],\;\delta=1,\;\mu_{t}\equiv 0.95$}

\hspace*{-0.4cm}\includegraphics[width=2.2in]{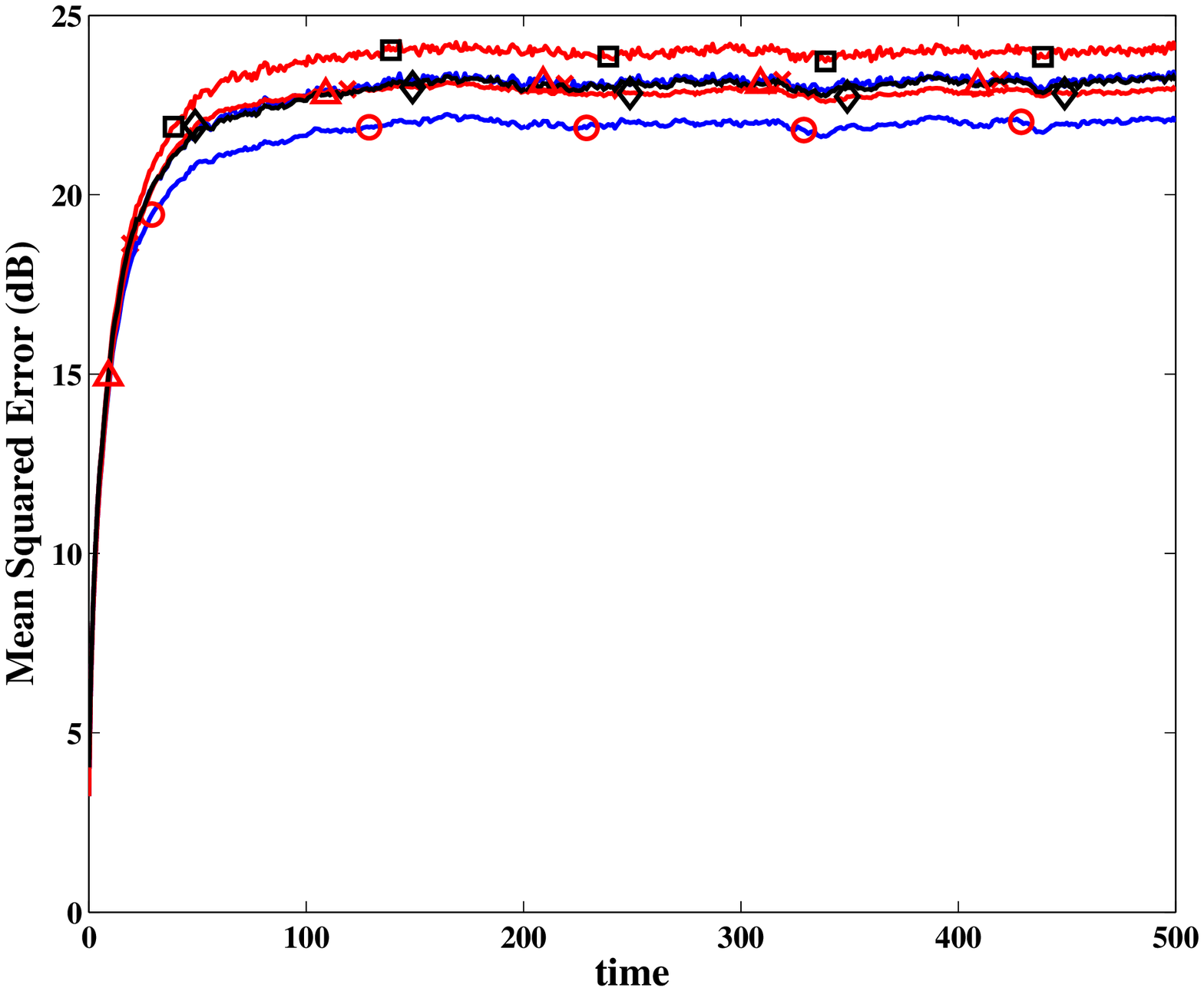}
\hspace{0.4cm}\includegraphics[width=2.2in]{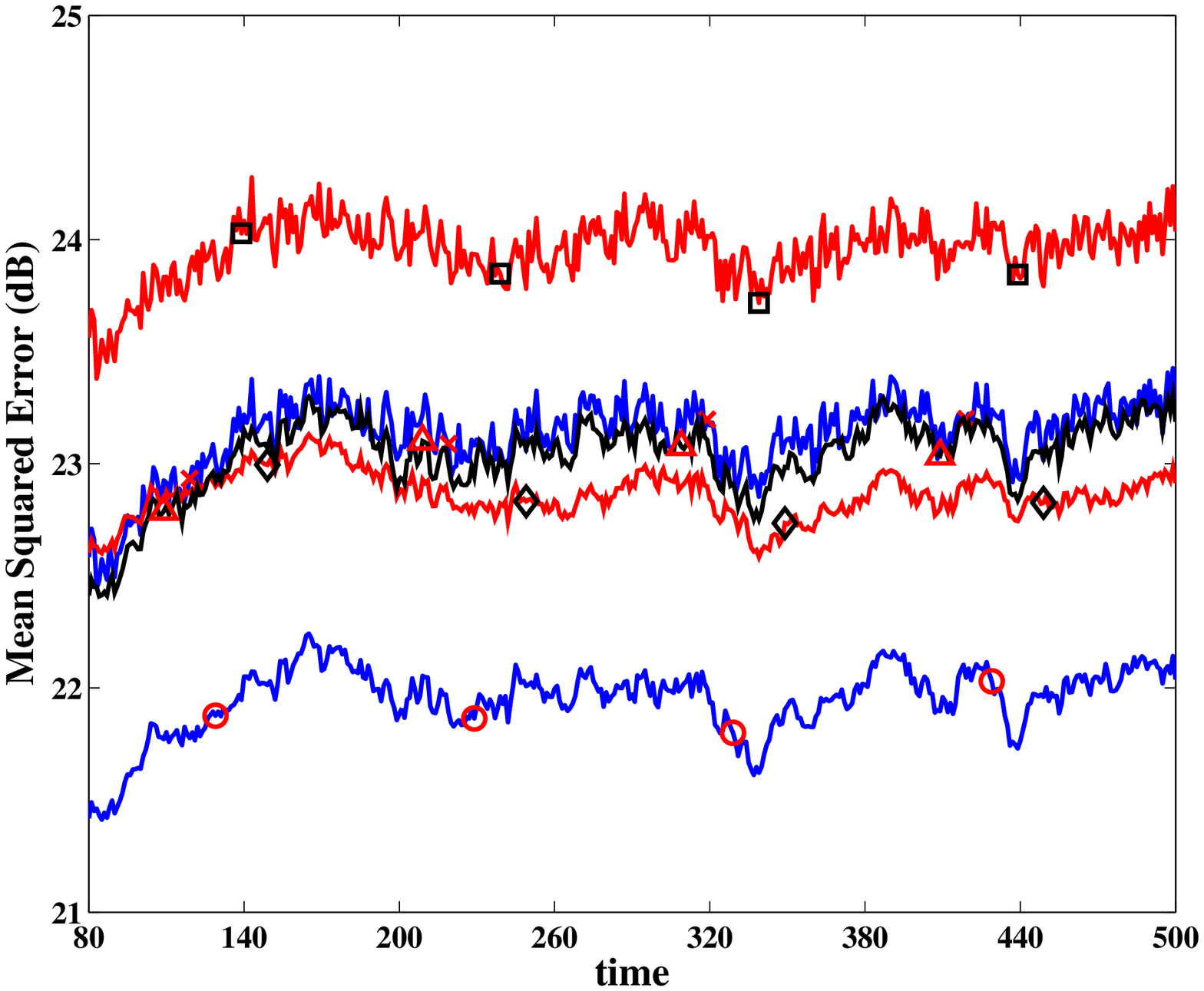}

\vspace{-0.2cm} \hspace*{-0.5cm} {\footnotesize (e)
$t\in[0,\;500],\;\delta=10,\;\mu_{t}\equiv 0.8$} \hspace{1.0cm}
{\footnotesize (f) $t\in[80,\;500],\;\delta=10,\;\mu_{t}\equiv 0.8$}
\end{center}

\vspace{-0.2cm} \caption{Empirical Mean Square Errors of
Estimations. $-\!\!\!-\!\!\!\!\!\Box\!\!\!\!\!-\!\!\!-$: Kalman
filter; $-\!\!\!-\!\!\!\!\!\Diamond\!\!\!\!\!-\!\!\!-$: estimator of
\cite{ssfpjs04}; $-\!\!\!-\!\!\!\!\!{\rm X}\!\!\!\!\!-\!\!\!-$:
estimator of \cite{zhou10b} ;
$-\!\!\!-\!\!\!\!\!\triangle\!\!\!\!\!-\!\!\!-$: estimator of
\cite{lz11}; $-\!\!\!-\!\!\!\!\!{\scriptsize
\bigcirc}\!\!\!\!\!-\!\!\!-$: estimator of this paper.}
\end{figure}

Kalman Filter, KFIO of \cite{ssfpjs04}, RSE of \cite{zhou10b}, the
RSE with missing measurements (RSEMM) developed in \cite{lz11}, as
well as the RSE developed in this paper (RSEIO), are utilized to
estimate the plant states. When the Kalman filter,
RSE of \cite{zhou10b} and RSEMM are utilized, every received $y_{t}$
is regarded as a plant output measurement. Empirical MSE is used to
measure estimation accuracy of these methods. More precisely,
$5\times 10^{3}$ numerical experiments are performed with the
temporal variable $t$ varies from $0$ to $5\times 10^{2}$. Let
$x_{t}^{[j]}$ and $\hat{x}_{t}^{[j]}$ represent respectively the
actual plant state and its estimate at the time instant $t$ in the
$j$-th numerical experiment. Then, the empirical
MSE of estimations at this time instant is defined as follows
\begin{displaymath}
\frac{1}{5\times 10^{3}}\sum_{j=1}^{5\times
10^{3}}[x_{t}^{[j]}-\hat{x}_{t}^{[j]}]^{T}[x_{t}^{[j]}-\hat{x}_{t}^{[j]}]
\end{displaymath}

In Figure 1a, simulation results with $\delta=1$ is shown. This case
is completely the same as that of \cite{lz11}. To make the
differences among these curves clear when the temporal variable $t$
takes a large value, in Figure 1b, they are re-plotted for the time
interval $80\leq t\leq 5\times 10^{2}$. From these simulations, it
becomes clear that when modelling errors fall into the interval
$[-1,\; 1]$, KFIO outperforms RSEIO. This is not a surprise, but
only means that for this numerical example, estimation accuracy of
the Kalman filter is not very sensitive to modelling errors, and in
order to make a better trade-off between nominal performance and
accuracy deteriorations, a greater value should be selected for the
design parameter $\mu_{t}$ of RSEIO. As a matter of fact, actual
computations show that if this design parameter is selected to be
$0.95$, then, RSEIO will have a slightly higher estimation accuracy
than KFIO. The corresponding results are given in
Figures 1c and 1d.

To clarify necessities to take into account of modelling errors in
state estimations, as well as influences of the design parameter
$\mu_{t}$ on estimation accuracy, simulation results with
$\delta=10$ and $\mu_{t}\equiv 0.8$ are also provided in Figures 1e
and 1f. Results of these sub-figures clearly show that when the
magnitude of modelling errors is large, sensitivity reduction for
the innovation process of the Kalman filter is really very helpful
in increasing its robustness against parametric modelling errors,
and therefore improve its estimation accuracy. It
is also clear from these simulation results that an appropriate
selection of the estimator design parameter $\mu_{t}$ heavily
depends on specific descriptions of modelling errors, such as their
variation intervals, etc.

In all these computations, RSEIO has a better estimation accuracy
than both RSE of \cite{zhou10b} and RSEMM of \cite{lz11}. This
result may imply that information about random measurement droppings
is more efficiently utilized by the estimation procedure of this
paper, and the cost function $J(x_{t|t+1},\;w_{t|t+1})$ of Equation
(\ref{eqn:7}) is more physically reasonable than that adopted in
\cite{lz11} when information is contained in $y_{t+1}$ about whether
or not it is a plant output measurement.

These simulation results also show that KFIO outperforms the
traditional Kalman filter appreciably, but in comparison with RSE of
\cite{zhou10b}, accuracy improvement by RSEMM is not very
significant.

\begin{figure}[!ht]
\begin{center}
\hspace*{-0.4cm}\includegraphics[width=2.2in]{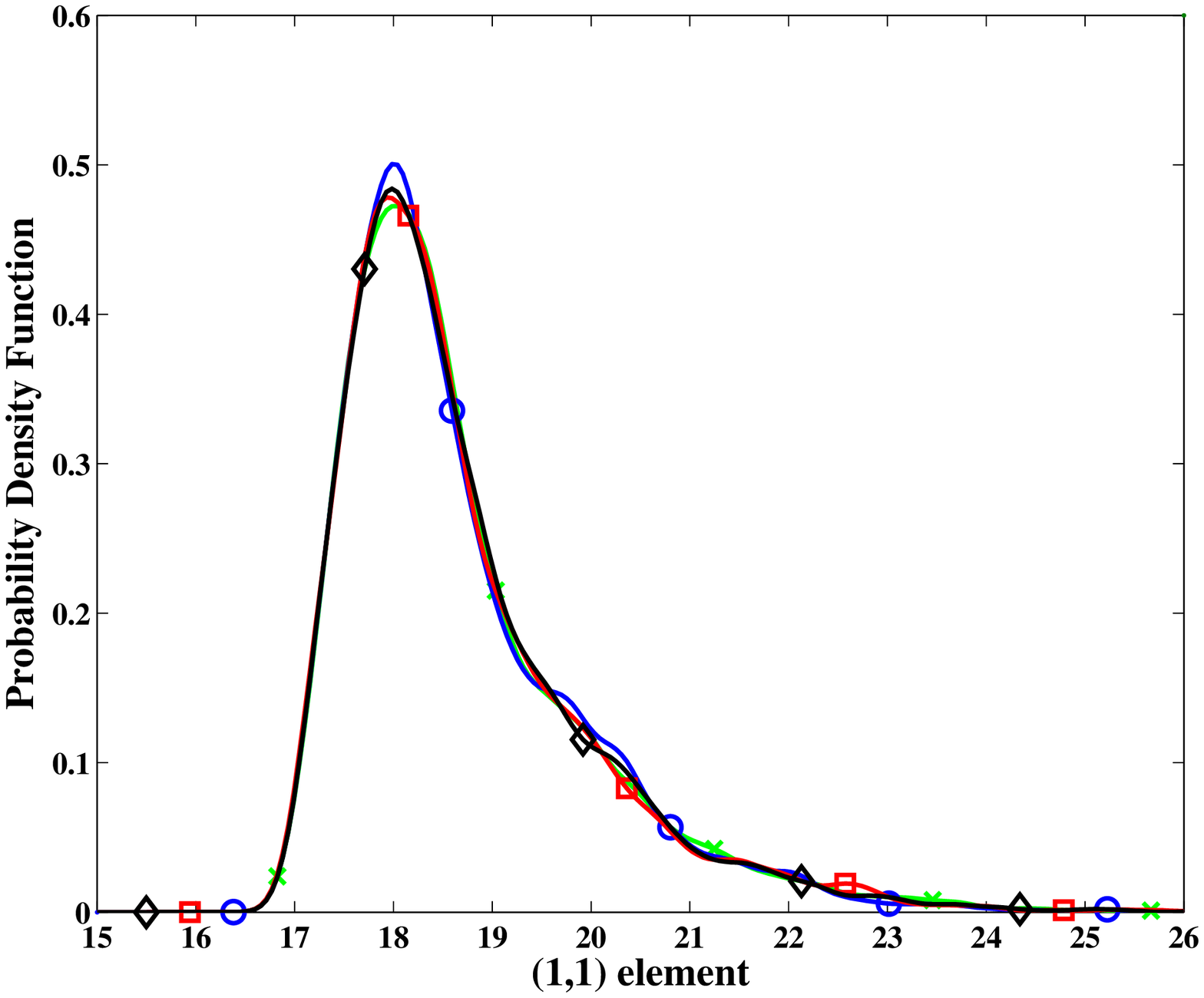}
\hspace{1.4cm}\includegraphics[width=2.2in]{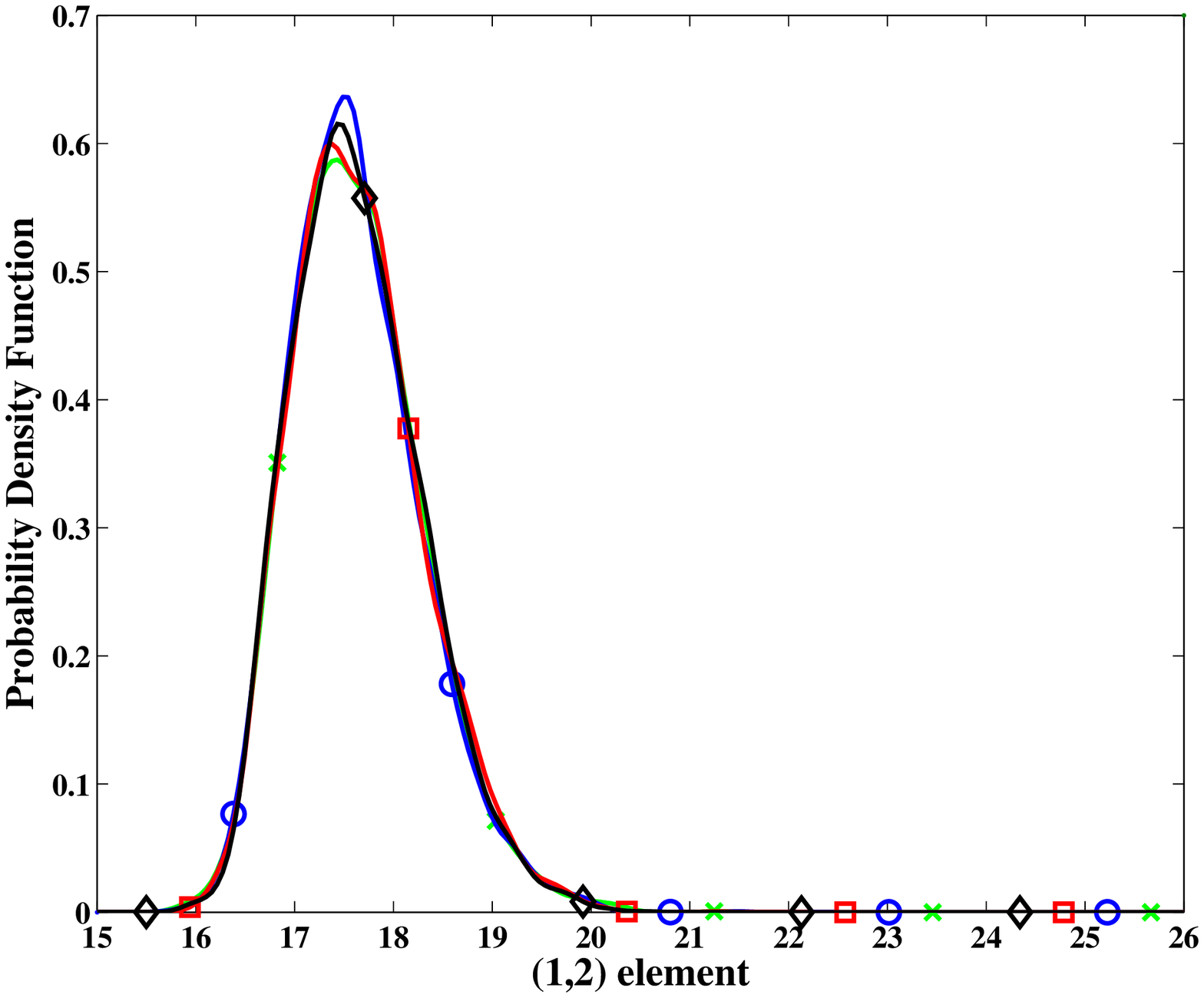}

\vspace{-0.3cm} \hspace*{-0.4cm} {\footnotesize (a) the 1st row 1st
column element} \hspace{1.9cm} {\footnotesize (b) the 1st row 2nd
column element}

\vspace{-0.0cm}
\hspace*{-0.4cm}\includegraphics[width=2.2in]{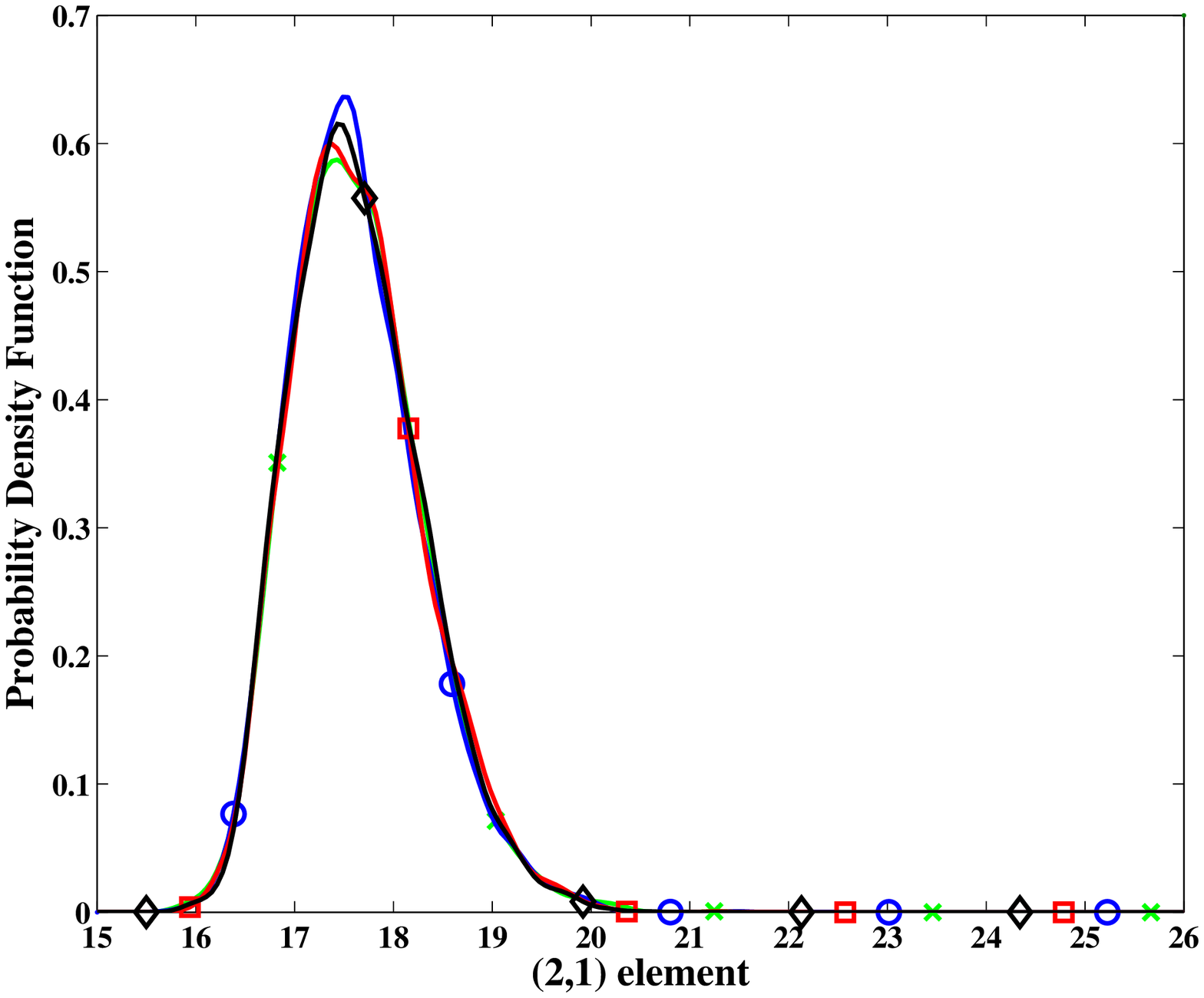}
\hspace{1.4cm}\includegraphics[width=2.2in]{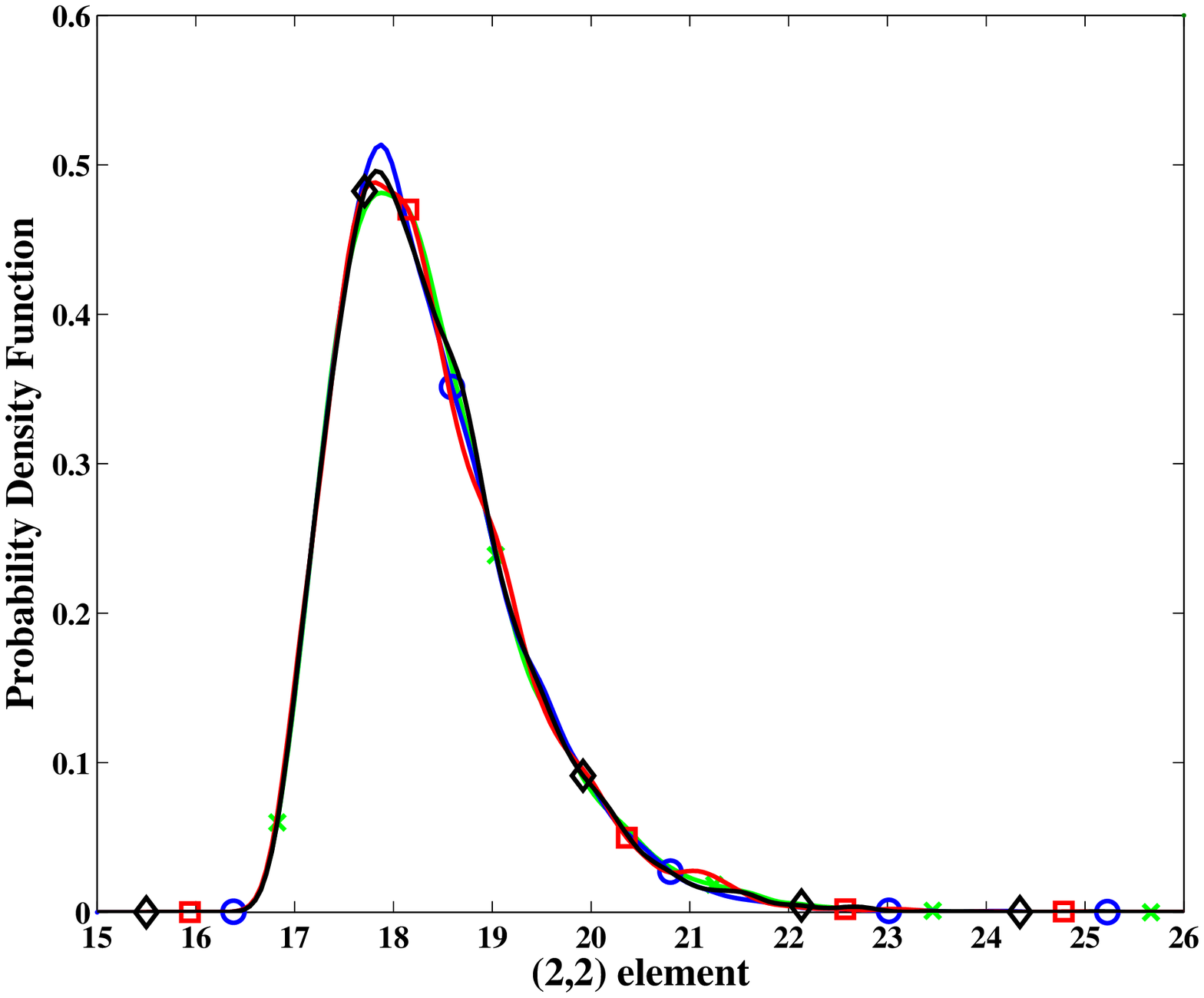}

\vspace{-0.3cm} \hspace*{-0.4cm} {\footnotesize (c) the 2nd row 1st
column element} \hspace{1.9cm} {\footnotesize (d) the 2nd row 2nd
column element}
\end{center}

\vspace{-0.2cm} \caption{Empirical Probability Density Function for
Elements of the Pseudo-Covariance Matrix at $t=500$.
$-\!\!\!-\!\!\!\!\!{\rm X}\!\!\!\!\!-\!\!\!-$: $P_{0|0}=0.1I_{2}$;
$-\!\!\!-\!\!\!\!\!{\scriptsize \bigcirc}\!\!\!\!\!-\!\!\!-$:
$P_{0|0}=I_{2}$; $-\!\!\!-\!\!\!\!\!\Box\!\!\!\!\!-\!\!\!-$:
$P_{0|0}=10I_{2}$; $-\!\!\!-\!\!\!\!\!\Diamond\!\!\!\!\!-\!\!\!-$:
$P_{0|0}=100I_{2}$. }
\end{figure}

In Figure 2, empirical probability density function
(EPDF) is shown for every element of the PCM $P_{t|t}$ at $t=5\times
10^{2}$ with 4 different initial $P_{0|0}$. In computing these
EPDFs, $5\times 10^{3}$ independent numerical experiments have been
performed for each situation and the Matlab file {\it ksdensity.m}
is used with default parameters in estimating the EPDF. Moreover,
the magnitude bound of modelling errors and the RSEIO design
parameter are respectively selected as $\delta=10$ and
$\mu_{t}\equiv 0.8$. From this figure, it is clear that although the
initial $P_{0|0}$s are significantly distinct from each other, the
EPDFs are very close for every element of the final $P_{500|500}$.
This confirms the theoretical results on the convergence of the
RESIO. On the other hand, it appears that the PDF of every element
of the stationary PCM is a continuous function, which is greatly
different from the conclusion about KFIO, in which it has been
demonstrated in \cite{km12} that the stationary distribution has a
fractured support. Moreover, the EPDFs of the non-diagonal elements
are almost the same. This is due to the symmetry of the PCM.

It is worthwhile to point out that the comparisons of \cite{lz11},
in both its theoretical analyzes and its numerical simulations, are
not appropriate, noting that in \cite{wyhl05}, a one-step recursive
robust state predictor is derived, while the problem discussed in
\cite{lz11} is to robustly estimate plant state using current and
past observations. In fact, from Figure 1, it is clear that
estimation accuracy of RSEMM is even slightly worse than the
traditional Kalman filter, in which neither parametric errors nor
random measurement droppings are taken into account\footnote[1]{When
the number of experiments is selected to be the same as that of
\cite{lz11}, that is, $5\times 10^{2}$, consistent observations have
been found, although the corresponding computation results fluctuate
more wildly.}. However, it is declared in \cite{lz11} that RSEMM is
slightly better than the estimator of \cite{wyhl05}, while
\cite{wyhl05} claims its superiority over the traditional Kalman
filter in prediction accuracies. These conclusions are apparently
contradictory. Moreover, the numerical examples adopted in these two
papers are completely different. In addition, time averaging is
adopted in \cite{wyhl05} for estimation accuracy evaluations, but
\cite{lz11} used ensemble averaging. These differences make the
comparisons more unreasonable and the conclusions more confusing.
Regretfully, these important things have been overlooked by this
author.

\section{Concluding Remarks}

In this paper, the sensitivity penalization based robust state
estimation procedure is extended to situations in which plant output
measurements may be randomly dropped due to communication failures.
A new recursion formula has been derived for the PCM of estimation
errors. Necessary and sufficient conditions have been established
for the strict contractiveness of an iteration of this recursion. It
has been proved that under some controllability and observability
conditions, as well as some weak restrictions on the arrival
probability of plant output measurements, the gain matrix of the
developed RSE converges with probability one to a stationary
distribution. Numerical simulations show that this RSE may
outperform the well known Kalman filter in estimation accuracy.

While some progress have been made in robust state estimations with
random measurement droppings, various important issues ask for
further efforts. Among them, more general and less conservative
conditions for the convergence of the obtained RSE, explicit
expressions for the stationary distribution of the PCM, etc., seem
essential in determining required capacity of a communication
channel and selecting a suitable estimator design parameter.

\renewcommand{\theequation}{a.\arabic{equation}}
\setcounter{equation}{0}

\section*{Appendix: Proof of Some Technical Results}

In order to prove the theoretical results of this paper, the
following results are required, which are well known in matrix
analysis and linear estimations, and can be straightforwardly proved
through algebraic manipulations \cite{ksh00,hj91}.

\hspace*{-0.4cm}{\bf Lemma A1.} For arbitrary matrices
$A,\;B,\;C,\;D$ with compatible dimensions, assume that all the
involved matrix inverses exist. Then
\begin{eqnarray}
& & \left[\begin{array}{cc} A & B \\ C & D \end{array}\right]=
\left[\begin{array}{cc} I & 0 \\ CA^{-1} & I \end{array}\right]
\left[\begin{array}{cc} A & 0 \\ 0 & D-CA^{-1}B \end{array}\right]
\left[\begin{array}{cc} I & A^{-1}B \\ 0 & I
\end{array}\right]\nonumber\\
& & \hspace*{2cm}= \left[\begin{array}{cc} I & BD^{-1} \\ 0 & I
\end{array}\right] \left[\begin{array}{cc} A-BD^{-1}C & 0 \\ 0 & D
\end{array}\right]
\left[\begin{array}{cc} I & 0 \\ D^{-1}C & I \end{array}\right] \\
& & [A+CBD]^{-1}=A^{-1}-A^{-1}C[B^{-1}+DA^{-1}C]^{-1}DA^{-1} \\
& & A(I+BA)^{-1}=(I+AB)^{-1}A
\end{eqnarray}

\hspace*{-0.4cm}{\bf Proof of Theorem 1:} For brevity, define
vectors $\alpha_{t}$ and $\alpha_{t0}$ respectively as
$\alpha_{t}={\rm\bf col}\!\{x_{t|t+1},\;w_{t|t+1}\}$ and
$\alpha_{t0}={\rm\bf col}\!\{\hat{x}_{t|t},\;0\}$. Moreover, define
matrices $\bar{P}_{t|t}$, $\bar{Q}_{t}$, $\bar{B}_{t}(0)$ and
$\bar{A}_{t}(0)$ respectively as
\begin{eqnarray*}
& &
\hspace*{-1cm}\bar{P}_{t|t}=(P_{t|t}^{-1}+\lambda_{t}\gamma_{t+1}S_{t}^{T}S_{t})^{-1},
\hspace{0.5cm}
\bar{Q}_{t}=\left[Q_{t}^{-1}+\lambda_{t}\gamma_{t+1}T_{t}^{T}(I+\lambda_{t}\gamma_{t+1}S_{t}P_{t|t}S_{t}^{T})T_{t}\right]^{-1} \\
& &\hspace*{-1cm}
\bar{B}_{t}(0)=B_{t}(0)-\lambda_{t}\gamma_{t+1}A_{t}(0)\bar{P}_{t|t}S_{t}^{T}T_{t},\hspace{0.25cm}
\bar{A}_{t}(0)=[A_{t}(0)-\lambda_{t}\gamma_{t+1}\bar{B}_{t}(0)\bar{Q}_{t}T_{t}^{T}S_{t}][I-\lambda_{t}\gamma_{t+1}\bar{P}_{t|t}S_{t}^{T}S_{t}]
\end{eqnarray*}
Furthermore, abbreviate $A_{t}(0)$, $\bar{A}_{t}(0)$, $B_{t}(0)$,
$\bar{B}_{t}(0)$ and $C_{t}(0)$ respectively as $A_{t}$,
$\bar{A}_{t}$, $B_{t}$, $\bar{B}_{t}$ and $C_{t}$. Note that for
every $k\in\{1,2,\cdots,n\}$, we have
\begin{eqnarray}
& & \frac{\partial
e_{t}(\varepsilon_{t},\;\varepsilon_{t+1})}{\partial
\varepsilon_{t,k}} =-C_{t+1}(\varepsilon_{t+1})\frac{\partial
A_{t}(\varepsilon_{t})}{\partial \varepsilon_{t,k}}
x_{t|t+1}-C_{t+1}(\varepsilon_{t+1})\frac{\partial
B_{t}(\varepsilon_{t})}{\partial
\varepsilon_{t,k}}w_{t|t+1} \\
& & \frac{\partial
e_{t}(\varepsilon_{t},\;\varepsilon_{t+1})}{\partial
\varepsilon_{t+1,k}} =-\frac{\partial
C_{t+1}(\varepsilon_{t+1})}{\partial \varepsilon_{t+1,k}}
A_{t}(\varepsilon_{t})x_{t|t+1}-\frac{\partial
C_{t+1}(\varepsilon_{t+1})}{\partial
\varepsilon_{t+1,k}}B_{t}(\varepsilon_{t})w_{t|t+1}
\end{eqnarray}
Then, from the definition of the cost function
$J(x_{t|t+1},w_{t|t+1})$, it can be straightforwardly proved that
\begin{equation}
J(\alpha_{t})\!=\!\frac{\mu_{t}}{2}\!\left\{\!\!(\star)^{T}\!\!{\rm\bf
diag}\!\left\{\!P_{t|t}^{-1}\!,\;\!Q_{t}^{-1}
\!\right\}\!\!(\alpha_{t}\!\!-\!\!\alpha_{t0})\!\!+\!\!\gamma_{t\!+\!1}(\star)^{T}R_{t\!+\!1}^{-1}(C_{t\!+\!1}[A_{t}\;
B_{t}]\alpha_{t}\!\!-\!\!y_{t+1})\!\!+\!\!\lambda_{t}\gamma_{t\!+\!1}(\star)^{T}([S_{t}\;
T_{t}]\alpha_{t})\!\!\right\}
\end{equation}
Therefore,
\begin{eqnarray}
\frac{\partial
J(\alpha_{t})}{\partial\alpha_{t}}\!\!\!\!&=&\!\!\!\!\mu_{t}\!\left\{\!\!{\rm\bf
diag}\left\{\! P_{t|t}^{-1},\; Q_{t}^{-1}
\!\right\}\!\!(\alpha_{t}\!-\!\alpha_{t0})\!+\!\gamma_{t+1}(C_{t+1}[A_{t}\;
B_{t}])^{T}R_{t+1}^{-1}(C_{t+1}[A_{t}\;
B_{t}]\alpha_{t}\!-\!y_{t+1})\!+\!\right.\nonumber\\
& &
\hspace*{8.5cm}\left.\lambda_{t}\gamma_{t+1}[S_{t}\;T_{t}]^{T}[S_{t}\;
T_{t}]\alpha_{t}\!\!\right\}\nonumber\\
&=&\!\!\!\!\mu_{t}\!\left\{\!\!\left(\!\!{\rm\bf diag}\!\left\{\!
P_{t|t}^{-1}\!,\;\! Q_{t}^{-1}
\!\right\}+\lambda_{t}\gamma_{t\!+\!1}\![S_{t}\;T_{t}]^{T}[S_{t}\;
T_{t}]\!+\!\gamma_{t\!+\!1}[A_{t}\;B_{t}]^{T}C_{t\!+\!1}^{T}R_{t\!+\!1}^{-1}C_{t\!+\!1}[A_{t}\;
B_{t}]\!\!\right)\!\!\alpha_{t}\!-\right.\nonumber\\
& & \hspace*{4cm}\left. {\rm\bf
diag}\left\{\!P_{t|t}^{-1},\;Q_{t}^{-1}
\!\right\}\alpha_{t0}-\gamma_{t+1}[A_{t}\;B_{t}]^{T}C_{t+1}^{T}R_{t+1}^{-1}y_{t+1}\right\}
\end{eqnarray}

Note that $J(\alpha_{t})$ is a convex function and $\mu_{t}\neq 0$.
It is obvious that the optimal $\alpha_{t}$, denote it by
$\hat{\alpha}_{t}$, which minimizes $J(\alpha_{t})$, is given by its
first derivative condition. That is,
\begin{eqnarray}
\hat{\alpha}_{t}\!\!&=&\!\!\left\{\!\!{\rm\bf
diag}\left\{\!P_{t|t}^{-1},\;\! Q_{t}^{-1}
\!\right\}+\lambda_{t}\gamma_{t+1}[S_{t}\;T_{t}]^{T}[S_{t}\;
T_{t}]+\gamma_{t+1}[A_{t}\;B_{t}]^{T}C_{t+1}^{T}R_{t+1}^{-1}C_{t+1}[A_{t}\;
B_{t}]\!\!\right\}^{-1}\times\nonumber\\
& & \hspace*{3cm}\left\{{\rm\bf diag}\left\{\!P_{t|t}^{-1},\;\!
Q_{t}^{-1}
\!\right\}\alpha_{t0}+\gamma_{t+1}[A_{t}\;B_{t}]^{T}C_{t+1}^{T}R_{t+1}^{-1}y_{t+1}\right\}
\label{eqn:a1}
\end{eqnarray}

On the other hand, direct algebraic manipulations show that
\begin{equation}
T_{t}^{T}T_{t}-\lambda_{t}\gamma_{t+1}T_{t}^{T}S_{t}[P_{t|t}^{-1}+\lambda_{t}\gamma_{t+1}S_{t}^{T}S_{t}]^{-1}S_{t}^{T}T_{t}=
T_{t}^{T}[I+\lambda_{t}\gamma_{t+1}S_{t}P_{t|t}S_{t}^{T}]^{-1}T_{t}
\end{equation}
Then, from Lemma A1 and the definitions of the matrices
$\bar{P}_{t|t}$ and $\bar{Q}_{t}$, the following relation can be
immediately obtained,
\begin{equation}
{\rm\bf diag}\left\{\!P_{t|t}^{-1},\;\! Q_{t}^{-1}
\!\right\}\!+\!\lambda_{t}\gamma_{t\!+\!1}[S_{t}\;T_{t}]^{T}[S_{t}\;
T_{t}]\!\!=\!\!\left[\!\!\begin{array}{cc} I & 0 \\
\lambda_{t}\gamma_{t\!+\!1}T_{t}^{T}S_{t}\bar{P}_{t|t} & I
\end{array}\!\!\right]\!
\left[\!\!\begin{array}{cc} \bar{P}_{t|t}^{-1} & 0 \\ 0 &
\bar{Q}_{t}^{-1}
\end{array}\!\!\right]
\left[\!\!\begin{array}{cc} I &
\lambda_{t}\gamma_{t\!+\!1}\bar{P}_{t|t}S_{t}^{T}T_{t} \\ 0 & I
\end{array}\!\!\right] \label{eqn:a2}
\end{equation}

Substitute this relation into Equation (\ref{eqn:a1}), it can be
further proved that
\begin{eqnarray}
\hat{\alpha}_{t}\!\!&=&\!\!\left\{\!\!\left[\begin{array}{cc} I & 0 \\
\lambda_{t}\gamma_{t+1}T_{t}^{T}S_{t}\bar{P}_{t|t} & I
\end{array}\right]
\left[\begin{array}{cc} \bar{P}_{t|t}^{-1} & 0 \\ 0 &
\bar{Q}_{t}^{-1}
\end{array}\right]
\left[\begin{array}{cc} I &
\lambda_{t}\gamma_{t+1}\bar{P}_{t|t}S_{t}^{T}T_{t} \\ 0 & I
\end{array}\right]+\right.\nonumber\\
& &
\hspace*{0.5cm}\left.\gamma_{t+1}[A_{t}\;B_{t}]^{T}C_{t+1}^{T}R_{t+1}^{-1}C_{t+1}[A_{t}\;
B_{t}]\!\!\right\}^{-1}\left\{{\rm\bf diag}\left\{\!P_{t|t}^{-1},\;
Q_{t}^{-1}\!\right\}\alpha_{t0}+\gamma_{t+1}[A_{t}\;B_{t}]^{T}C_{t+1}^{T}R_{t+1}^{-1}y_{t+1}\right\}\nonumber\\
&=& \!\!\left[\begin{array}{cc} I &
-\lambda_{t}\gamma_{t+1}\bar{P}_{t|t}S_{t}^{T}T_{t} \\ 0 & I
\end{array}\right]\left\{\!\!\left[\begin{array}{cc} \bar{P}_{t|t}^{-1} & 0 \\ 0 &
\bar{Q}_{t}^{-1}
\end{array}\right]+\gamma_{t+1}[A_{t}\;\bar{B}_{t}]^{T}C_{t+1}^{T}R_{t+1}^{-1}C_{t+1}[A_{t}\;
\bar{B}_{t}]\!\!\right\}^{-1}\times\nonumber\\
& & \hspace*{3cm}\left[{\rm\bf col}\!\!\left\{I,\;\!
-\lambda_{t}\gamma_{t+1}T_{t}^{T}S_{t}\bar{P}_{t|t}\right\}P_{t|t}^{-1}\hat{x}_{t|t}+\gamma_{t+1}[A_{t}\;\bar{B}_{t}]^{T}C_{t+1}^{T}R_{t+1}^{-1}y_{t+1}\right]
\end{eqnarray}

Hence,
\begin{eqnarray}
\hat{x}_{t+1|t+1}\!\!&=&\!\![A_{t}\;B_{t}]\hat{\alpha}_{t} \nonumber\\
&=&\!\! [A_{t}\;\bar{B}_{t}]\left\{\!\!{\rm\bf
diag}\left\{\!\bar{P}_{t|t}^{-1},\; \bar{Q}_{t}^{-1}
\right\}+\gamma_{t+1}[A_{t}\;\bar{B}_{t}]^{T}C_{t+1}^{T}R_{t+1}^{-1}C_{t+1}[A_{t}\;
\bar{B}_{t}]\!\!\right\}^{-1}\times\nonumber\\
& & \hspace*{3cm}\left\{{\rm\bf col}\!\!\left\{I,\;\!
-\lambda_{t}\gamma_{t+1}T_{t}^{T}S_{t}\bar{P}_{t|t}\right\}P_{t|t}^{-1}\hat{x}_{t|t}+\gamma_{t+1}[A_{t}\;\bar{B}_{t}]^{T}C_{t+1}^{T}R_{t+1}^{-1}y_{t+1}\right\}
\nonumber\\
&=&\!\! [A_{t}\;\bar{B}_{t}]\left\{\!\!I+\gamma_{t+1}{\rm\bf
diag}\left\{\!\bar{P}_{t|t},\;
\bar{Q}_{t}\right\}[A_{t}\;\bar{B}_{t}]^{T}C_{t+1}^{T}R_{t+1}^{-1}C_{t+1}[A_{t}\;
\bar{B}_{t}]\!\right\}^{-1}{\rm\bf
diag}\left\{\!\bar{P}_{t|t},\;\bar{Q}_{t}
\!\right\}\times\nonumber\\
& & \hspace*{3cm}\left\{{\rm\bf col}\!\!\left\{I,\;\!
-\lambda_{t}\gamma_{t+1}T_{t}^{T}S_{t}\bar{P}_{t|t}\right\}P_{t|t}^{-1}\hat{x}_{t|t}+\gamma_{t+1}[A_{t}\;\bar{B}_{t}]^{T}C_{t+1}^{T}R_{t+1}^{-1}y_{t+1}\right\}\nonumber\\
&=&\!\! \left\{\!\!I+\gamma_{t+1}[A_{t}\;\bar{B}_{t}]{\rm\bf
diag}\left\{\!\bar{P}_{t|t},\;\bar{Q}_{t}
\!\right\}[A_{t}\;\bar{B}_{t}]^{T}C_{t+1}^{T}R_{t+1}^{-1}C_{t+1}\!\!\right\}^{-1}[A_{t}\;\bar{B}_{t}]{\rm\bf
diag}\left\{\!\bar{P}_{t|t},\;\bar{Q}_{t}
\!\right\}\times\nonumber\\
& & \hspace*{3cm}\left\{{\rm\bf col}\!\!\left\{I,\;\!
-\lambda_{t}\gamma_{t+1}T_{t}^{T}S_{t}\bar{P}_{t|t}\right\}P_{t|t}^{-1}\hat{x}_{t|t}+\gamma_{t+1}[A_{t}\;\bar{B}_{t}]^{T}C_{t+1}^{T}R_{t+1}^{-1}y_{t+1}\right\}\nonumber\\
&=&\!\!\left[I+\gamma_{t+1}P_{t+1|t}C_{t+1}^{T}R_{t+1}^{-1}C_{t+1}\right]^{-1}\left\{\bar{A}_{t}\hat{x}_{t|t}+\gamma_{t+1}P_{t+1|t}C_{t+1}^{T}R_{t+1}^{-1}y_{t+1}\right\}
\end{eqnarray}
in which
$P_{t+1|t}=A_{t}\bar{P}_{t|t}A_{t}^{T}+\bar{B}_{t}\bar{Q}_{t}\bar{B}_{t}^{T}$.
In the derivation of the last equality of the above equation, the
relation
$\bar{P}_{t|t}P_{t|t}^{-1}=I-\lambda_{t}\gamma_{t+1}\bar{P}_{t|t}S_{t}^{T}S_{t}$
has been utilized, which is a direct result of the definition of the
matrix $\bar{P}_{t|t}$.

Therefore,
\begin{eqnarray}
\hat{x}_{t+1|t+1}\!\!&=&\!\!\bar{A}_{t}\hat{x}_{t|t}+\gamma_{t+1}\left[I+\gamma_{t+1}P_{t+1|t}C_{t+1}^{T}R_{t+1}^{-1}C_{t+1}\right]^{-1}P_{t+1|t}C_{t+1}^{T}R_{t+1}^{-1}y_{t+1}\nonumber\\
& & \hspace*{2cm}
-\gamma_{t+1}\left[I+\gamma_{t+1}P_{t+1|t}C_{t+1}^{T}R_{t+1}^{-1}C_{t+1}\right]^{-1}P_{t+1|t}C_{t+1}^{T}R_{t+1}^{-1}C_{t+1}\bar{A}_{t}\hat{x}_{t|t}\nonumber\\
&=&\!\!\bar{A}_{t}\hat{x}_{t|t}+\gamma_{t+1}\left[P_{t+1|t}^{-1}+\gamma_{t+1}C_{t+1}^{T}R_{t+1}^{-1}C_{t+1}\right]^{-1}C_{t+1}^{T}R_{t+1}^{-1}\left\{y_{t+1}
-C_{t+1}\bar{A}_{t}\hat{x}_{t|t}\right\}
\end{eqnarray}

Comparing this recursive formula for
$\hat{x}_{t+1|t+1}$ with that of the Kalman filter given in
\cite{kalman60,ksh00,simon06}, it is clear that the matrix
$[P_{t+1|t}^{-1}+\gamma_{t+1}C_{t+1}^{T}R_{t+1}^{-1}C_{t+1}]^{-1}$
plays the same role as that of the covariance matrix of estimation
errors in Kalman filtering. It is therefore reasonable to denote it
by $P_{t+1|t+1}$. The proof can now be completed by noting that if
$\gamma_{t+1}=0$, then $\bar{A}_{t}=A_{t}(0)$,
$\bar{B}_{t}=B_{t}(0)$, $\bar{P}_{t|t}=P_{t|t}$ and
$\bar{Q}_{t}=Q_{t}$, as well as that if $\gamma_{t+1}=1$, then
$\bar{A}_{t}=\hat{A}_{t}(0)$, $\bar{B}_{t}=\hat{B}_{t}(0)$,
$\bar{P}_{t|t}=\hat{P}_{t|t}$ and $\bar{Q}_{t}=\hat{Q}_{t}$.
\hspace{\fill}$\Diamond$

\hspace*{-0.4cm}{\bf Proof of Theorem 2:} To simplify mathematical
expressions, in this proof, $A_{t}(0)$, $B_{t}(0)$ and $C_{t+1}(0)$
are again respectively abbreviated to be $A_{t}$, $B_{t}$ and
$C_{t+1}$. From the proof of Theorem 1, it is clear that when
$\gamma_{t+1}=1$,
\begin{eqnarray}
\hat{x}_{t+1|t+1}\!\!&=&\!\![A_{t}\;B_{t}]\left\{\!\!{\rm\bf
diag}\left\{\!P_{t|t}^{-1},\;\! Q_{t}^{-1}
\!\right\}+\lambda_{t}[S_{t}\;T_{t}]^{T}[S_{t}\;
T_{t}]+[A_{t}\;B_{t}]^{T}C_{t+1}^{T}R_{t+1}^{-1}C_{t+1}[A_{t}\;
B_{t}]\!\!\right\}^{-1}\times\nonumber\\
& & \hspace*{3cm}\left\{{\rm\bf
diag}\!\!\left\{\!P_{t|t}^{-1},\;Q_{t}^{-1}\!\right\}\!{\rm\bf
col}\!\!\left\{\hat{x}_{t|t},\;
0\right\}+[A_{t}\;B_{t}]^{T}C_{t+1}^{T}R_{t+1}^{-1}y_{t+1}\right\}
\label{eqn:a4}
\end{eqnarray}
Moreover, Theorem 1 also declares that under such a situation,
\begin{equation}
\hat{x}_{t+1|t+1}=\hat{A}_{t}\hat{x}_{t|t}+P_{t+1|t+1}C_{t+1}^{T}R_{t+1}^{-1}[y_{t+1}-C_{t+1}\hat{A}_{t}\hat{x}_{t|t}]
\label{eqn:a5}
\end{equation}

As Equations (\ref{eqn:a4}) and (\ref{eqn:a5}) are just two
different expressions for the same state estimate
$\hat{x}_{t+1|t+1}$, the coefficient matrices respectively for
$\hat{x}_{t|t}$ and $y_{t+1}$ should be equal to each other. A
comparison of the coefficient matrices of $y_{t+1}$ show that
\begin{equation}
P_{t+1|t+1}=[A_{t}\;B_{t}]\left\{\!\!{\rm\bf
diag}\left\{\!P_{t|t}^{-1},\;\! Q_{t}^{-1}
\!\right\}+\lambda_{t}[S_{t}\;T_{t}]^{T}[S_{t}\;
T_{t}]+[A_{t}\;B_{t}]^{T}C_{t+1}^{T}R_{t+1}^{-1}C_{t+1}[A_{t}\;
B_{t}]\!\!\right\}^{\!-1}\!\!\!\!\![A_{t}\;B_{t}]^{T} \label{eqn:a6}
\end{equation}

On the other hand, direct algebraic operations show that
\begin{eqnarray}
\lambda_{t}S_{t}^{T}S_{t}-\lambda_{t}^{2}S_{t}^{T}T_{t}[Q_{t}^{-1}+\lambda_{t}T_{t}^{T}T_{t}]^{-1}T_{t}^{T}S_{t}
&=&\lambda_{t}S_{t}^{T}\left\{I-\lambda_{t}T_{t}[I+\lambda_{t}Q_{t}T_{t}^{T}T_{t}]^{-1}Q_{t}T_{t}^{T}\right\}S_{t}\nonumber\\
&=&\lambda_{t}S_{t}^{T}[I+\lambda_{t}T_{t}Q_{t}T_{t}^{T}]^{-1}S_{t}\nonumber\\
&=&\tilde{S}_{t}^{T}\tilde{S}_{t}
\end{eqnarray}
Then, from Lemma A1 and the definition of $\check{Q}_{t}$, the
following relation can be immediately obtained,
\begin{equation}
{\rm\bf diag}\left\{\!P_{t|t}^{-1},\;\! Q_{t}^{-1}
\!\right\}\!+\!\lambda_{t}[S_{t}\;T_{t}]^{T}[S_{t}\; T_{t}]\!=\!
\left[\!\!\begin{array}{cc} I &
\lambda_{t}S_{t}^{T}T_{t}\check{Q}_{t}
\\ 0 & I
\end{array}\!\!\right]\!
\left[\!\!\begin{array}{cc} P_{t|t}^{-1}+\tilde{S}_{t}^{T}\tilde{S}_{t} & 0 \\
0 & \check{Q}_{t}^{-1}
\end{array}\!\!\right]\!
\left[\!\!\begin{array}{cc} I & 0 \\
\lambda_{t}\check{Q}_{t}T_{t}^{T}S_{t} & I
\end{array}\!\!\right] \label{eqn:a3}
\end{equation}

Substitute Equation (\ref{eqn:a3}) into Equation (\ref{eqn:a6}), we
have
\begin{eqnarray}
P_{t+1|t+1}\!\!&=&\!\!\!\!\left(\!\![A_{t}\;B_{t}]
\left[\!\!\begin{array}{cc} I & 0 \\
\lambda_{t}\check{Q}_{t}T_{t}^{T}S_{t} & I
\end{array}\!\!\right]^{\!\!-1}\!\right)\!\!\left\{\!\!\left[\!\!\begin{array}{cc}
P_{t|t}^{-1}\!+\!\tilde{S}_{t}^{T}\tilde{S}_{t} & 0
\\ 0 & \check{Q}_{t}^{-1}
\end{array}\!\!\right]\!+\!\left(\!\![A_{t}\;B_{t}]
\left[\begin{array}{cc} I & 0 \\
\lambda_{t}\check{Q}_{t}T_{t}^{T}S_{t} & I
\end{array}\right]^{\!\!-1}\!\right)^{T}\right.\!\!\!\!\times\nonumber\\
& &
\hspace*{0.5cm}\left.C_{t+1}^{T}R_{t+1}^{-1}C_{t+1}\left([A_{t}\;
B_{t}]\!\!\left[\!\!\begin{array}{cc} I & 0 \\
\lambda_{t}\check{Q}_{t}T_{t}^{T}S_{t} & I
\end{array}\!\!\right]^{-1}\!\right)\!\!\right\}^{\!-1}\!\!\!\!\left(\!\![A_{t}\;B_{t}]
\!\!\left[\!\!\begin{array}{cc} I & 0 \\
\lambda_{t}\check{Q}_{t}T_{t}^{T}S_{t} & I
\end{array}\!\!\right]^{-1}\!\right)^{T} \nonumber\\
&=&\!\!\!\![\check{A}_{t}\;B_{t}]\!\!\left\{\!\!{\rm\bf
diag}\left\{\!
P_{t|t}^{-1}\!+\!\tilde{S}_{t}^{T}\tilde{S}_{t},\;\check{Q}_{t}^{-1}
\!\right\}\!+\![\check{A}_{t}\;B_{t}]^{T}C_{t+1}^{T}R_{t+1}^{-1}C_{t+1}
[\check{A}_{t}\;B_{t}]\!\!\right\}^{-1}\!\![\check{A}_{t}\;B_{t}]^{T}\nonumber\\
&=&\!\!\!\!\left\{\!\!I\!+\![\check{A}_{t}\;B_{t}]{\rm\bf
diag}\left\{\!
(P_{t|t}^{-1}\!+\!\tilde{S}_{t}^{T}\tilde{S}_{t})^{-1},\;\check{Q}_{t}
\!\right\}\!\![\check{A}_{t}\;B_{t}]^{T}C_{t+1}^{T}R_{t+1}^{-1}C_{t+1}\!\!\right\}^{\!\!-1}\!\!\!\!\times\nonumber\\
& & \hspace*{6cm}[\check{A}_{t}\;B_{t}]{\rm\bf diag}\left\{\!
(P_{t|t}^{-1}\!+\!\tilde{S}_{t}^{T}\tilde{S}_{t})^{-1},\;
\check{Q}_{t}\!\right\}\!\![\check{A}_{t}\;B_{t}]^{T}\nonumber\\
&=&\!\!\!\!\left\{\left[\check{A}_{t}(P_{t|t}^{-1}\!+\!\tilde{S}_{t}^{T}\tilde{S}_{t})^{-1}
\check{A}^{T}_{t}+B_{t}\check{Q}_{t}B_{t}^{T}\right]^{-1}+C_{t+1}^{T}R_{t+1}C_{t+1}\right\}^{-1}
\label{eqn:a7}
\end{eqnarray}

When $\check{A}_{t}$ is invertible, from the definition of the
matrix $\tilde{B}_{t}$, we have that
\begin{equation}
\check{A}_{t}(P_{t|t}^{-1}\!+\!\tilde{S}_{t}^{T}\tilde{S}_{t})^{-1}
\check{A}^{T}_{t}+B_{t}\check{Q}_{t}B_{t}^{T}=\check{A}_{t}\left\{(P_{t|t}^{-1}\!+\!\tilde{S}_{t}^{T}\tilde{S}_{t})^{-1}
+\tilde{B}_{t}\check{Q}_{t}\tilde{B}_{t}^{T}\right\}\check{A}^{T}_{t}
\label{eqn:a8}
\end{equation}

Note that
\begin{eqnarray}
& & \left\{(P_{t|t}^{-1}\!+\!\tilde{S}_{t}^{T}\tilde{S}_{t})^{-1}
+\tilde{B}_{t}\check{Q}_{t}\tilde{B}_{t}^{T}\right\}^{-1}\nonumber\\
&=&\!\!\!\!\left\{I+(P_{t|t}^{-1}\!+\!\tilde{S}_{t}^{T}\tilde{S}_{t})\tilde{B}_{t}\check{Q}_{t}\tilde{B}_{t}^{T}\right\}^{-1}
(P_{t|t}^{-1}\!+\!\tilde{S}_{t}^{T}\tilde{S}_{t}) \nonumber\\
&=&\!\!\!\!\left\{\tilde{B}_{t}\check{Q}_{t}\tilde{B}_{t}^{T}+P_{t|t}(I+\tilde{S}_{t}^{T}\tilde{S}_{t}\tilde{B}_{t}\check{Q}_{t}\tilde{B}_{t}^{T})\right\}^{-1}(I+P_{t|t}
\tilde{S}_{t}^{T}\tilde{S}_{t})\nonumber\\
&=&\!\!\!\!\left\{\tilde{B}_{t}\check{Q}_{t}\tilde{B}_{t}^{T}+P_{t|t}(I+\tilde{S}_{t}^{T}\tilde{S}_{t}\tilde{B}_{t}\check{Q}_{t}\tilde{B}_{t}^{T})\right\}^{-1}
\left\{I+\left[\tilde{B}_{t}\check{Q}_{t}\tilde{B}_{t}^{T}+P_{t|t}(I+\tilde{S}_{t}^{T}\tilde{S}_{t}\tilde{B}_{t}\check{Q}_{t}\tilde{B}_{t}^{T})\right]
(I+\right.\nonumber\\
&
&\hspace{4cm}\left.\tilde{S}_{t}^{T}\tilde{S}_{t}\tilde{B}_{t}\check{Q}_{t}\tilde{B}_{t}^{T})^{-1}\tilde{S}_{t}^{T}\tilde{S}_{t}-\tilde{B}_{t}\check{Q}_{t}\tilde{B}_{t}^{T}
(I+\tilde{S}_{t}^{T}\tilde{S}_{t}\tilde{B}_{t}\check{Q}_{t}\tilde{B}_{t}^{T})^{-1}\tilde{S}_{t}^{T}\tilde{S}_{t}\right\}\nonumber\\
&=&\!\!\!\!\tilde{S}_{t}^{T}(I+\tilde{S}_{t}\tilde{B}_{t}\check{Q}_{t}\tilde{B}_{t}^{T}\tilde{S}_{t}^{T})^{-1}\tilde{S}_{t}+
\left\{\tilde{B}_{t}(\check{Q}_{t}+\check{Q}_{t}\tilde{B}_{t}^{T}\tilde{S}_{t}^{T}\tilde{S}_{t}\tilde{B}_{t}\check{Q}_{t})\tilde{B}_{t}^{T}
+(I+\tilde{B}_{t}\check{Q}_{t}\tilde{B}_{t}^{T}\tilde{S}_{t}^{T}\tilde{S}_{t})P_{t|t}\!\times\right.\nonumber\\
& & \hspace{8cm}\left.
(I+\tilde{B}_{t}\check{Q}_{t}\tilde{B}_{t}^{T}\tilde{S}_{t}^{T}\tilde{S}_{t})^{T}\right\}^{-1}
\label{eqn:a9}
\end{eqnarray}

Substitute Equations (\ref{eqn:a8}) and (\ref{eqn:a9}) into Equation
(\ref{eqn:a7}), the following recursive expression for $P_{t+1|t+1}$
is obtained for situations when $\check{A}_{t}$ is invertible,
\begin{eqnarray}
P_{t+1|t+1}^{-1}\!\!\!\!&=&\!\!\!\!\check{A}_{t}^{-T}\left[(P_{t|t}^{-1}\!+\!\tilde{S}_{t}^{T}\tilde{S}_{t})^{-1}
+\tilde{B}_{t}\check{Q}_{t}\tilde{B}_{t}^{T}\right]^{-1}\check{A}_{t}^{-1}+C_{t+1}^{T}R_{t+1}C_{t+1}\nonumber\\
&=&\!\!\!\! \check{A}_{t}^{-T}\!\!
\left\{\!\!\tilde{B}_{t}(\check{Q}_{t}\!+\!\check{Q}_{t}\tilde{B}_{t}^{T}\tilde{S}_{t}^{T}\tilde{S}_{t}\tilde{B}_{t}\check{Q}_{t})\tilde{B}_{t}^{T}
\!+\!(I\!+\!\tilde{B}_{t}\check{Q}_{t}\tilde{B}_{t}^{T}\tilde{S}_{t}^{T}\tilde{S}_{t})P_{t|t}
(I\!+\!\tilde{B}_{t}\check{Q}_{t}\tilde{B}_{t}^{T}\tilde{S}_{t}^{T}\tilde{S}_{t})^{T}\!\right\}^{\!-1}\!\!
\check{A}_{t}^{\!-1}\!+\nonumber\\
& & \hspace*{4cm}
\check{A}_{t}^{-T}\tilde{S}_{t}^{T}(I+\tilde{S}_{t}\tilde{B}_{t}\check{Q}_{t}\tilde{B}_{t}^{T}\tilde{S}_{t}^{T})^{-1}\tilde{S}_{t}
\check{A}_{t}^{-1}+C_{t+1}^{T}R_{t+1}C_{t+1}\nonumber\\
&=&\!\!\!\!\left\{\!\!B_{t}(\check{Q}_{t}\!+\!\check{Q}_{t}\tilde{B}_{t}^{T}\tilde{S}_{t}^{T}\tilde{S}_{t}\tilde{B}_{t}\check{Q}_{t})B_{t}^{T}
\!+\!(\check{A}_{t}\!+\!B_{t}\check{Q}_{t}\tilde{B}_{t}^{T}\tilde{S}_{t}^{T}\tilde{S}_{t})P_{t|t}
(\check{A}_{t}\!+\!B_{t}\check{Q}_{t}\tilde{B}_{t}^{T}\tilde{S}_{t}^{T}\tilde{S}_{t})^{T}\!\right\}^{\!-1}\!\!\!+\nonumber\\
& & \hspace*{4cm}
(\tilde{S}_{t}\check{A}_{t}^{-1})^{T}(I+\tilde{S}_{t}\tilde{B}_{t}\check{Q}_{t}\tilde{B}_{t}^{T}\tilde{S}_{t}^{T})^{-1}(\tilde{S}_{t}
\check{A}_{t}^{-1})+C_{t+1}^{T}R_{t+1}C_{t+1}\nonumber\\
&=&\!\!\!\![\tilde{A}_{t}P_{t|t}\tilde{A}_{t}^{T}+B_{t}\tilde{Q}_{t}B_{t}^{T}]^{-1}+\tilde{C}_{t+1}^{T}\tilde{R}_{t+1}^{-1}\tilde{C}_{t+1}
\end{eqnarray}

This completes the proof. \hspace{\fill}$\Diamond$

\hspace*{-0.4cm}{\bf Proof of Theorem 3:} Define matrix $G_{o}$ as
\begin{displaymath}
G_{o}=\Phi_{N,11}^{T}\Phi_{N,21}+\sum_{i=N-1}^{1}\left[\left(\prod_{k=N}^{i}
\Phi_{k,11}^{T}\right)\Phi_{i,21}\left(\prod_{k=i+1}^{N}\Phi_{k,11}\right)\right]
\end{displaymath}
Then, it can be claimed from Lemma 2 that
$\prod_{t=1}^{N}\Phi_{t}\in {\cal H}_{l}$ if and only if ${\rm\bf
D}_{et}\left\{G_{o} \right\}\neq 0$.

Assume that at the sampling instants $t_{1},\;t_{2},\; \cdots\,\;
t_{p}$, with $1\leq t_{1}<t_{2}<\cdots<t_{p}\leq N$, $\gamma_{k}=1$;
and at any other sampling instants between $1$ and $N$,
$\gamma_{k}=0$. Define $t_{0}$ as $t_{0}=0$. Then, according to the
definition of $\Phi_{t}$, the following relation is obtained,
\begin{equation}
G_{o}\!=\!\sum_{j=0}^{p}\!\left[\!\left(\!\prod_{k=t_{j}+1}^{N}
\Phi_{k,11}\!\right)^{T}\!\Phi_{t_{j},11}^{T}\Phi_{t_{j},21}\!\left(\!\prod_{k=t_{j}+1}^{N}\Phi_{k,11}\!\right)\!\right]
\label{eqn:a10}
\end{equation}
in which $\prod_{k=q}^{N}\Phi_{k,11}$ is defined to be the identity
matrix if $q>N$. This situation occurs when $t_{p}=N$.

Note that for every $t_{j}$ with $1\leq j\leq p$,
\begin{equation}
\Phi_{t_{j},11}^{T}\Phi_{t_{j},12}=A^{[1]T}H^{[1]T}H^{[1]}A^{[1]}
=(H^{[1]}A^{[1]})^{T}(H^{[1]}A^{[1]}) \label{eqn:a11}
\end{equation}
Moreover,
\begin{eqnarray}
\prod_{k=t_{j}+1}^{N}\Phi_{k,11}&=&\prod_{k=t_{j}+1}^{t_{j+1}-1}\Phi_{k,11}
\times\Phi_{t_{j+1},11}\times
\prod_{k=t_{j+1}+1}^{t_{j+2}-1}\Phi_{k,11}\times\Phi_{t_{j+2},11}\times\cdots\times
\prod_{k=t_{p}+1}^{N}\Phi_{k,11} \nonumber\\
&=&
(A^{[2]})^{t_{j+1}-t_{j}-1}A^{[1]}(A^{[2]})^{t_{j+2}-t_{j+1}-1}A^{[1]}\cdots
(A^{[2]})^{N-t_{p}} \nonumber\\
&=&
\left(\prod_{s=j}^{p-1}\left[(A^{[2]})^{t_{s+1}-t_{s}-1}A^{[1]}\right]\right)
(A^{[2]})^{N-t_{p}} \label{eqn:a12}
\end{eqnarray}

Substitute Equations (\ref{eqn:a11}) and (\ref{eqn:a12}) into
Equation (\ref{eqn:a10}), the following relation is obtained,
\begin{eqnarray}
G_{o}\!\!\!\!&=&\!\!\!\!\sum_{j=0}^{p}\left\{\!\!\left[\star\right]^{T}
(H^{[1]}A^{[1]})^{T}(H^{[1]}A^{[1]})
\left[\!\!\left(\prod_{s=j}^{p-1}\left[(A^{[2]})^{t_{s+1}-t_{s}-1}A^{[1]}\right]\right)
(A^{[2]})^{N-t_{p}}\right)\right\}\nonumber\\
&=&\!\!\!\!\sum_{j=0}^{p}\left\{\!\!\left[\star\right]^{T}
\left[H^{[1]}A^{[1]}\left(\prod_{s=j}^{p-1}\left[(A^{[2]})^{t_{s+1}-t_{s}-1}A^{[1]}\right]\right)
(A^{[2]})^{N-t_{p}}\right)\right\}\nonumber\\
&=&\!\!\!\!\left[\star\right]^{T}\left[\begin{array}{c} H^{[1]}(A^{[2]})^{N-t_{p}} \\
H^{[1]}A^{[1]}(A^{[2]})^{t_{p}-t_{p-1}-1}(A^{[2]})^{N-t_{p}} \\
\vdots \\
H^{[1]}\prod_{j=1}^{p}\left(A^{[1]}(A^{[2]})^{t_{j}-t_{j-1}-1}\right)(A^{[2]})^{N-t_{p}}
\end{array}\right]\nonumber\\
&=&\!\!\!\! (A^{[2]T})^{N-t_{p}}O_{b}^{T}O_{b}(A^{[2]})^{N-t_{p}}
\end{eqnarray}

Recall that the matrix $A^{[2]}$ is assumed invertible. It is
obvious from the above equality that the satisfaction of the
inequality ${\rm\bf D}_{et}\left\{G_{o} \right\}\neq 0$ is
equivalent to that the matrix $O_{b}$ is of full column rank. This
completes the proof. \hspace{\fill}$\Diamond$

\hspace*{-0.4cm}{\bf Proof of Theorem 4:} From Lemma 2, it can be
claimed that $\prod_{t=1}^{N}\Phi_{t}\in {\cal H}_{r}$ if and only
if ${\rm\bf D}_{et}\left\{G_{c}\right\}\neq 0$, in which
\begin{displaymath}
G_{c}=\Phi_{1,12}\Phi_{1,11}^{T}+\sum_{i=2}^{N}\left[\left(\prod_{k=1}^{i-1}
\Phi_{k,11}\right)\Phi_{i,12}\left(\prod_{k=i}^{1}\Phi_{k,11}^{T}\right)\right]
\end{displaymath}

Similar to the proof of Theorem 3, assume that at the sampling
instants $t_{1},\;t_{2},\; \cdots\,\; t_{p}$, with $1\leq
t_{1}<t_{2}<\cdots<t_{p}\leq N$, $\gamma_{k}=1$; and at any other
sampling instants between $1$ and $N$, $\gamma_{k}=0$. Moreover,
$t_{0}$ is once again defined as $t_{0}=0$. Furthermore, define
$t_{p+1}$ as $t_{p+1}=N+1$.

Define $\prod_{k=1}^{0}\Phi_{k,11}$ as the identity matrix. It can
then be easily seen that,
\begin{equation}
G_{c}=\sum_{i=1}^{N}\left[\left(\prod_{k=1}^{i-1}
\Phi_{k,11}\right)\Phi_{i,12}\Phi_{i,11}^{T}\left(\prod_{k=1}^{i-1}\Phi_{k,11}\right)^{T}\right]
\label{eqn:a13}
\end{equation}

When $i\in\{t_{1},\;t_{2},\;\cdots,\; t_{p}\}$, assume that
$i=t_{j}$, $j=1,2,\cdots,p$. We have
\begin{eqnarray}
\Phi_{i,12}\Phi_{i,11}^{T}\!\!\!\!&=&\!\!\!\!\left[G^{[1]}G^{[1]T}(A^{[1]})^{-T}\right]A^{[1]T}=G^{[1]}G^{[1]T}
\\
\prod_{k=1}^{i-1}\Phi_{k,11}\!\!\!\!&=&\!\!\!\!\left(\prod_{k=1}^{t_{1}-1}\Phi_{k,11}\right)\Phi_{t_{1},11}
\left(\prod_{k=t_{1}+1}^{t_{2}-1}\Phi_{k,11}\right)\Phi_{t_{2},11}\cdots
\left(\prod_{k=t_{j-1}+1}^{t_{j}-1}\Phi_{k,11}\right)\nonumber\\
&=&\!\!\!\!(A^{[2]})^{t_{1}-1}A^{[1]}(A^{[2]})^{t_{2}-t_{1}-1}A^{[1]}\cdots
(A^{[2]})^{t_{j}-t_{j-1}-1}\nonumber\\
&=&\!\!\!\!(A^{[2]})^{t_{1}-1}\prod_{s=1}^{j-1}\left[A^{[1]}(A^{[2]})^{t_{s+1}-t_{s}-1}\right]
\label{eqn:a15}
\end{eqnarray}

When $i\not\in\{t_{1},\;t_{2},\;\cdots,\; t_{p}\}$, assume that
$t_{j-1}<i<t_{j}$, $j=1,2,\cdots,p+1$. On the basis of Equation
(\ref{eqn:a15}), we then have
\begin{eqnarray}
\Phi_{i,12}\Phi_{i,11}^{T}\!\!\!\!&=&\!\!\!\!\left[G^{[2]}G^{[2]T}(A^{[2]})^{-T}\right]A^{[2]T}=G^{[2]}G^{[2]T}
\\
\prod_{k=1}^{i-1}\Phi_{k,11}\!\!\!\!&=&\!\!\!\!\left(\prod_{k=1}^{t_{j-1}-1}\Phi_{k,11}\right)\Phi_{t_{j-1},11}
\left(\prod_{k=t_{j-1}+1}^{i-1}\Phi_{k,11}\right)\nonumber\\
&=&\!\!\!\!(A^{[2]})^{t_{1}-1}\left(\prod_{s=1}^{j-2}\left[A^{[1]}(A^{[2]})^{t_{s+1}-t_{s}-1}\right]\right)
A^{[1]}(A^{[2]})^{i-t_{j-1}-1}
\end{eqnarray}

When $1\leq i<t_{1}$, direct algebraic manipulations show that
\begin{eqnarray}
\Phi_{i,12}\Phi_{i,11}^{T}\!\!\!\!&=&\!\!\!\!\left[G^{[2]}G^{[2]T}(A^{[2]})^{-T}\right]A^{[2]T}=G^{[2]}G^{[2]T}
\\
\prod_{k=1}^{i-1}\Phi_{k,11}\!\!\!\!&=&\!\!\!\!\prod_{k=1}^{i-1}A^{[2]}\!=\!\left(A^{[2]}\right)^{i-1}
\end{eqnarray}

Substitute these relations into Equation (\ref{eqn:a13}), the
following equalities are obtained.
\begin{eqnarray}
G_{c}\!\!\!\!&=&\!\!\!\!\sum_{i=1}^{t_{1}-1}\left[\left(\prod_{k=1}^{i-1}
\Phi_{k,11}\right)\Phi_{i,12}\Phi_{i,11}^{T}\left(\star\right)^{T}\right]+
\sum_{j=2}^{p+1}\left\{\left[\left(\prod_{k=1}^{t_{j-1}-1}
\Phi_{k,11}\right)\Phi_{t_{j-1},12}\Phi_{t_{j-1},11}^{T}\left(\star\right)^{T}\right]+\right.\nonumber\\
& &
\hspace*{6cm}\left.\sum_{i=t_{j-1}+1}^{t_{j}-1}\left[\left(\prod_{k=1}^{i-1}
\Phi_{k,11}\right)\Phi_{i,12}\Phi_{i,11}^{T}\left(\star\right)^{T}\right]\right\}\nonumber\\
&=&\!\!\!\!\sum_{i=0}^{t_{1}-2}\left[\left(A^{[2]}\right)^{i}G^{[2]}\right]\left[\star\right]^{T}+
(A^{[2]})^{t_{1}-1}
\sum_{j=2}^{p+1}\left\{\left(\prod_{s=1}^{j-1}\left[A^{[1]}(A^{[2]})^{t_{s+1}-t_{s}-1}\right]
G^{[1]}\right)\left(\star\right)^{T}+\right.\nonumber\\
& &
\hspace*{0.5cm}\left.\sum_{i=t_{j-1}+1}^{t_{j}-1}\left(A^{[1]}\prod_{s=1}^{j-2}\left[(A^{[2]})^{t_{s+1}-t_{s}-1}A^{[1]}\right]
(A^{[2]})^{i-t_{j-1}-1}G^{[2]}\right)\left(\star\right)^{T}\right\}(A^{[2]T})^{t_{1}-1}\nonumber\\
&=&\!\!\!\!C_{o}C_{o}^{T}
\end{eqnarray}

Therefore, the inequality ${\rm\bf D}_{et}\left\{G_{c} \right\}\neq
0$ is satisfied, if and only if the matrix $C_{o}$ is of full row
rank. This completes the proof.  \hspace{\fill}$\Diamond$

\hspace*{-0.4cm}{\bf Proof of Lemma 3:} When
$\gamma_{k}|_{k=1}^{\infty}$ is white and has a Bernoulli
distribution, we have
\begin{eqnarray}
{\rm\bf
P}_{r}\left(\Gamma^{[N]}=S_{m}^{[N]}\right)&=&\prod_{k=N}^{1}{\rm\bf
P}_{r}\left(\gamma_{k}=S_{m}^{[N]}(k)\right)\nonumber \\
&=&
\prod_{k=N}^{1}\bar{\gamma}^{S_{m}^{[N]}(k)}(1-\bar{\gamma})^{1-S_{m}^{[N]}(k)}
\end{eqnarray}
Hence,
\begin{eqnarray}
{\rm log}\!\left[{\rm\bf
P}_{r}\left(\Gamma^{[N]}=S_{m}^{[N]}\right)\right]&=&\sum_{i=1}^{N}\left[S_{m}^{[N]}(k){\rm
log}\!(\bar{\gamma})+(1-S_{m}^{[N]}(k)){\rm
log}\!(1-\bar{\gamma})\right]\nonumber\\
&=&{\rm log}(\bar{\gamma})\sum_{i=1}^{N}S_{m}^{[N]}(i)+{\rm
log}(1-\bar{\gamma})\left(N-\sum_{i=1}^{N}S_{m}^{[N]}(i)\right)
\end{eqnarray}

When $\gamma_{k}|_{k=1}^{\infty}$ is a Markov chain,
\begin{equation}
{\rm\bf
P}_{r}\left(\Gamma^{[N]}=S_{m}^{[N]}\right)=\left\{\prod_{k=N}^{2}{\rm\bf
P}_{r}\left(\left.\gamma_{k}=S_{m}^{[N]}(k)\right|\gamma_{k-1}=S_{m}^{[N]}(k-1)\right)\right\}
{\rm\bf P}_{r}\left(\gamma_{1}=S_{m}^{[N]}(1)\right) \label{eqn:4}
\end{equation}

Note that
\begin{equation}
{\rm\bf
P}_{r}\left(\gamma_{1}=S_{m}^{[N]}(1)\right)=\left\{\begin{array}{ll}
(1-\alpha){\rm\bf P}_{r}(\gamma_{0}=1)+\beta {\rm\bf
P}_{r}(\gamma_{0}=0)   & S_{m}^{[N]}(1)=0 \\
\alpha{\rm\bf P}_{r}(\gamma_{0}=1)+(1-\beta){\rm\bf
P}_{r}(\gamma_{0}=0)   & S_{m}^{[N]}(1)=1 \end{array}\right.
\end{equation}
It can therefore be concluded from the assumption ${\rm\bf
P}_{r}(\gamma_{0}=1)=\bar{\gamma}$ and the fact that ${\rm\bf
P}_{r}(\gamma_{0}=0)+{\rm\bf P}_{r}(\gamma_{0}=1)\equiv 1$ that
\begin{eqnarray}
{\rm\bf P}_{r}\left(\gamma_{1}=S_{m}^{[N]}(1)\right)&=&
[1-S_{m}^{[N]}(1)]\left[(1-\alpha){\rm\bf P}_{r}(\gamma_{0}=1)+\beta
{\rm\bf
P}_{r}(\gamma_{0}=0)\right]+\nonumber\\
& & \hspace*{1cm}S_{m}^{[N]}(1)\left[\alpha{\rm\bf
P}_{r}(\gamma_{0}=1)+(1-\beta){\rm\bf
P}_{r}(\gamma_{0}=0)\right]\nonumber\\
&=&S_{m}^{[N]}(1)\!+\![1\!-\!2S_{m}^{[N]}(1)][\beta\!+\!\bar{\gamma}(1\!-\!\alpha\!-\!\beta)]
\label{eqn:5}
\end{eqnarray}

On the other hand, for an arbitrary $k\in\{2,\;3,\;\cdots,\; N\}$,
it can be straightforwardly declared from the definition of a Markov
chain and the assumption $0<\alpha,\;\beta<1$ that
\begin{eqnarray}
& & {\rm\bf
P}_{r}\left(\left.\gamma_{k}=S_{m}^{[N]}(k)\right|\gamma_{k-1}=S_{m}^{[N]}(k-1)\right)
\nonumber\\
&=& \left\{\begin{array}{lll} \alpha, & &
S_{m}^{[N]}(k)=1,\;S_{m}^{[N]}(k-1)=1 \\
1-\alpha, & &
S_{m}^{[N]}(k)=1,\;S_{m}^{[N]}(k-1)=0 \\
\beta, & &
S_{m}^{[N]}(k)=0,\;S_{m}^{[N]}(k-1)=1 \\
1-\beta, &\;\;\;\; & S_{m}^{[N]}(k)=0,\;S_{m}^{[N]}(k-1)=0 \end{array}\right.\nonumber\\
&=&\alpha^{S_{m}^{[N]}(k)S_{m}^{[N]}(k-1)}(1-\alpha)^{(1-S_{m}^{[N]}(k))S_{m}^{[N]}(k-1)}
(1-\beta)^{S_{m}^{[N]}(k)(1-S_{m}^{[N]}(k-1))}\beta^{(1-S_{m}^{[N]}(k))(1-S_{m}^{[N]}(k-1))}
\nonumber\\
&=&\beta\left(\frac{1-\alpha}{\beta}\right)^{S_{m}^{[N]}(k-1)}\left(\frac{1-\beta}{\beta}\right)^{S_{m}^{[N]}(k)}
\left(\frac{\alpha\beta}{(1-\alpha)(1-\beta)}\right)^{S_{m}^{[N]}(k)S_{m}^{[N]}(k-1)}
\label{eqn:6}
\end{eqnarray}

Substitute Equations (\ref{eqn:5}) and (\ref{eqn:6}) into Equation
(\ref{eqn:4}), the following relation is obtained,
\begin{eqnarray}
{\rm log}\!\left[{\rm\bf
P}_{r}\!\left(\!\Gamma^{[N]}\!=\!S_{m}^{[N]}\right)\!\right]\!\!\!\!&=&\!\!\!\!
\sum_{k=N}^{2}{\rm log}\!\left[{\rm\bf
P}_{r}\left(\left.\gamma_{k}=S_{m}^{[N]}(k)\right|\gamma_{k-1}=S_{m}^{[N]}(k-1)\right)\right]
+{\rm log}\!\left[{\rm\bf
P}_{r}\left(\gamma_{1}=S_{m}^{[N]}(1)\right)\right]\nonumber\\
&=&\!\!\!\! \sum_{k=N}^{2}\left\{{\rm log}(\beta)+
S_{m}^{[N]}(k-1){\rm log}\!\left(\frac{1-\alpha}{\beta}\right)
+S_{m}^{[N]}(k){\rm log}\!\left(\frac{1-\beta}{\beta}\right)
+\right.\nonumber\\
& & \hspace*{2cm}\left.S_{m}^{[N]}(k)S_{m}^{[N]}(k-1){\rm
log}\!\left(\frac{\alpha\beta}{(1-\alpha)(1-\beta)}\right)\right\}+\nonumber\\
& & \hspace*{2cm}{\rm
log}\!\left\{S_{m}^{[N]}(1)\!+\![1\!-\!2S_{m}^{[N]}(1)][\beta\!+\!\bar{\gamma}(1\!-\!\alpha\!-\!\beta)]\right\}\nonumber\\
&=&\!\!\!\!(N\!-\!1){\rm log}(\beta)\!+\!{\rm
log}\!\left(\!\frac{1\!-\!\alpha}{\beta}\!\right)\!\sum_{k=1}^{N-1}\!S_{m}^{[N]}(k)\!+\!{\rm
log}\!\left(\!\frac{1}{\beta}\!-\!1\!\right)\!\sum_{k=2}^{N}S_{m}^{[N]}(k)\!+\nonumber\\
& &\!\!\!\!\hspace*{2cm}{\rm
log}\!\left(\!\frac{\alpha\beta}{(1\!-\!\alpha)(1\!-\!\beta)}\!\right)\!\sum_{k=2}^{N}\!\left[\!S_{m}^{[N]}(k)S_{m}^{[N]}(k\!-\!1)\!\right]\!+\nonumber\\
& &\!\!\!\! \hspace*{2cm}{\rm
log}\!\left\{\!S_{m}^{[N]}(1)\!+\![1\!-\!2S_{m}^{[N]}(1)][\beta\!+\!\bar{\gamma}(1\!-\!\alpha\!-\!\beta)]\!\right\}
\end{eqnarray}

This completes the proof. \hspace{\fill}$\Diamond$

\hspace*{-0.4cm}{\bf Proof of Theorem 5:} Assume that there exist
two positive integers $m_{1}$ and $m_{2}$ such that the matrix pairs
$(A^{[1]}(A^{[2]})^{m_{1}},\;H^{[1]})$ and
$((A^{[2]})^{m_{2}}A^{[1]},\;G^{[2]})$ are respectively observable
and controllable. Then, the following two matrices $\bar{O}_{b}$ and
$\bar{C}_{n}$ are respectively of full column rank and full row
rank,
\begin{displaymath}
\bar{O}_{b}=\left[\begin{array}{c} H^{[1]} \\
H^{[1]}A^{[1]}(A^{[2]})^{m_{1}} \\ \vdots \\
H^{[1]}\left(A^{[1]}(A^{[2]})^{m_{1}}\right)^{n-1}
\end{array}\right],\hspace{0.5cm}
\bar{C}_{n}=\left[G^{[2]}\;\; (A^{[2]})^{m_{2}}A^{[1]}G^{[2]} \cdots
\left((A^{[2]})^{m_{2}}A^{[1]}\right)^{n-1}G^{[2]}\right]
\end{displaymath}

Define positive integers $N_{*}$ and $N$, as well as a finite binary
sequence $R^{[N]}=\{R^{[N]}(t)|_{t=1}^{N}\}$, respectively as
$N_{*}=(n-2)(m_{1}+1)+1$, $N=(n-2)(m_{1}+m_{2}+2)+2$, and
\begin{equation}
R^{[N]}(t)=\left\{\begin{array}{lll}
0 & t\in \left(1+(j-1)(m_{1}+1),\; 1+j(m_{1}+1)\right)\\
1 & t=1+(j-1)(m_{1}+1) \\
0 & t=N_{*}+(j-1)(m_{2}+1)+1\\
1 & t\in \left(N_{*}+(j-1)(m_{2}+1)+1,\;N_{*}+j(m_{2}+1)+1\right)
\end{array}\right.
\end{equation}
in which $j=1,2,\cdots,n-1$. Then, it can be claimed from Theorems 3
and 4 that when the finite random sequence $\Gamma^{[N]}$ has the
realization $R^{[N]}$, the corresponding matrices
$\Phi_{R^{[N]}(t)}|_{t=1}^{N}$ simultaneously satisfy
$\prod_{t=1}^{N_{*}}\Phi_{R^{[N]}(t)}\in{\cal H}_{l}$ and
$\prod_{t=N_{*}+1}^{N}\Phi_{R^{[N]}(t)}\in{\cal H}_{r}$. Hence,
according to Lemma 2,
$\prod_{t=1}^{N}\Phi_{R^{[N]}(t)}=\prod_{t=1}^{N_{*}}\Phi_{R^{[N]}(t)}\prod_{t=N_{*}+1}^{N}\Phi_{R^{[N]}(t)}$
belongs to both ${\cal H}_{l}$ and ${\cal H}_{r}$, and therefore
${\cal H}_{lr}$. It can therefore be declared from Lemma 2 that with
respect to this particular realization of $\gamma_{t}|_{t=1}^{N}$,
the corresponding matrix valued function ${\rm\bf
H}_{m}\left(\prod_{t=1}^{N}\Phi_{R^{[N]}(t)},\star\right)$, which is
defined on the set of $n\times n$ dimensional positive definite
matrices, is strictly contractive under the Riemannian distance
defined in Equation (\ref{eqn:10}). This means that if during the
time interval $[1,\;N]$, the random measurement dropping process
$\gamma_{t}|_{t=1}^{\infty}$ has this particular realization, then,
${\rm\bf L}_{ip}\!\left({\rm\bf
H}_{m}\!\left(\prod_{t=1}^{N}\Phi_{R^{[N]}(t)},\star\right)\!\right)$
$<1$. From Lemma 1, this inequality is further equivalent to
\begin{equation}
{\rm\bf L}_{ip}\left({\rm\bf H}_{m}\left(\Phi_{R^{[N]}(1)},{\rm\bf
H}_{m}\left(\Phi_{R^{[N]}(2)},\cdots,{\rm\bf
H}_{m}\left(\Phi_{R^{[N]}(N)},\star\right)\cdots\right)\right)\right)<1
\end{equation}

On the other hand, from the definition of the set ${\cal S}^{[N]}$,
it is obvious that $R^{[N]}\in {\cal S}^{[N]}$. Hence, according to
Lemma 4, when the measurement dropping process
$\gamma_{t}|_{t=1}^{\infty}$ has an independent and identical
Bernoulli distribution with a constant positive expectation, the
probability of the occurrence of this sequence is certainly greater
than $0$.

In addition, from Lemma 2 and the fact that $\Phi_{t}\in{\cal H}$ no
matter $\gamma_{t}=1$ or $\gamma_{t}=0$, it can be declared that for
every other element $S_{m}^{[N]}$ of the set ${\cal S}^{[N]}$, the
random alternative Lyapunov and Riccati recursions corresponding to
the particular realization $\Gamma^{[N]}=S_{m}^{[N]}$ of the
pseudo-covariance matrix $P_{t|t}$ in RSEIO, satisfies ${\rm\bf
L}_{ip}\left({\rm\bf
H}_{m}\left(\prod_{t=1}^{N}\Phi_{S_{m}^{[N]}(t)},\star\right)\right)\leq
1$, which is further equivalent to
\begin{equation}
{\rm\bf L}_{ip}\left({\rm\bf
H}_{m}\left(\Phi_{S_{m}^{[N]}(1)},{\rm\bf
H}_{m}\left(\Phi_{S_{m}^{[N]}(2)},\cdots,{\rm\bf
H}_{m}\left(\Phi_{S_{m}^{[N]}(N)},\star\right)\cdots\right)\right)\right)\leq
1
\end{equation}

Therefore,
\begin{eqnarray}
& & {\rm\bf E}_{\{\gamma_{t}|_{t=1}^{N}\}}\left\{{\rm log} {\rm\bf
L}_{ip}\!\left({\rm\bf H}_{m}\left(\Phi_{1},{\rm\bf
H}_{m}\left(\Phi_{2},\cdots,{\rm\bf
H}_{m}\left(\Phi_{N},\star\right)\cdots\right)\right)\right)\right\}\nonumber\\
&=& {\rm log}{\rm\bf L}_{ip}\!\left({\rm\bf
H}_{m}\left(\Phi_{R^{[N]}(1)},{\rm\bf
H}_{m}\left(\Phi_{R^{[N]}(2)},\cdots,{\rm\bf
H}_{m}\left(\Phi_{R^{[N]}(N)},\star\right)\cdots\right)\right)\right){\rm\bf
P}_{r}\left(\Gamma^{[N]}=R^{[N]}\right)+\nonumber \\
& & \sum_{S_{m}^{[N]}\in{\cal S}^{[N]}\backslash
R^{[N]}}\!\!\!\!\!\!\!\!\!\!\!\!{\rm log}{\rm\bf
L}_{ip}\!\left({\rm\bf H}_{m}\left(\Phi_{S_{m}^{[N]}(1)},{\rm\bf
H}_{m}\left(\Phi_{S_{m}^{[N]}(2)},\cdots,{\rm\bf
H}_{m}\left(\Phi_{S_{m}^{[N]}(N)},\star\right)\cdots\right)\right)\right){\rm\bf
P}_{r}\left(\Gamma^{[N]}=S_{m}^{[N]}\right)\nonumber\\
&<& 0
\end{eqnarray}

It can therefore be declared from Lemma 3 that, if the random
measurement dropping process $\gamma_{t}|_{t=1}^{\infty}$ has an
independent and identical Bernoulli distribution with a constant
positive expectation, and there exist two positive integers $m_{1}$
and $m_{2}$ such that the matrix pair
$(A^{[1]}(A^{[2]})^{m_{1}},\;H^{[1]})$ is observable and the matrix
pair $((A^{[2]})^{m_{2}}A^{[1]},\;G^{[2]})$ is controllable, then,
with the increment of the variable $t$, the pseudo-covariance matrix
$P_{t|t}$ of RSEIO converges with probability one to a stationary
distribution that is independent of its initial value $P_{0|0}$.

The other situations can be proved using completely similar
arguments. The details are therefore omitted.

This completes the proof. \hspace{\fill}$\Diamond$

\end{document}